\documentclass[fleqn,10pt,twocolumn]{wlscirep}
\usepackage[utf8]{inputenc}
\usepackage[T1]{fontenc}
\usepackage{bm}
\usepackage{amssymb}
\title{New physics from the polarised light of the cosmic microwave background}

\author[1,2,*]{Eiichiro Komatsu}
\affil[1]{Max-Planck-Institut f\"{u}r Astrophysik, Karl-Schwarzschild Str. 1, 85741 Garching, Germany}
\affil[2]{Kavli Institute for the Physics and Mathematics of the Universe (Kavli IPMU, WPI), University of Tokyo, Chiba 277-8582, Japan}

\affil[*]{e-mail:komatsu@mpa-garching.mpg.de}

\begin{abstract}
Cosmology requires new physics beyond the Standard Model of elementary particles and fields. What is the fundamental physics behind dark matter and dark energy? What generated the initial fluctuations in the early Universe? Polarised light of the cosmic microwave background (CMB) may hold the key to answers. In this article, we discuss two new developments in this research area. First, if the physics behind dark matter and dark energy violates parity symmetry, their coupling to photons rotates the plane of linear polarisation as the CMB photons travel more than 13 billion years. This effect is known as `cosmic birefringence': space filled with dark matter and dark energy behaves as if it were a birefringent material, like a crystal. A tantalising hint for such a signal has been found with the statistical significance of $3\sigma$. Next, the period of accelerated expansion in the very early Universe, called `cosmic inflation', produced a stochastic background of primordial gravitational waves (GW). What generated GW? The leading idea is vacuum fluctuations in spacetime, but matter fields could also produce a significant amplitude of primordial GW. Finding its origin using CMB polarisation opens a new window into the physics behind inflation. These new scientific targets may influence how data from future CMB experiments are collected, calibrated, and analysed.
\end{abstract}
\begin{document}

\flushbottom
\maketitle

\thispagestyle{empty}



The standard cosmological model, called $\Lambda$CDM, demands new physics beyond the Standard Model (SM) of elementary particles and fields\cite{weinberg:2008}. `$\Lambda$' denotes Einstein's cosmological constant, which is the simplest (but most difficult to understand\cite{weinberg:1989,martin:2012}) candidate of dark energy responsible for the accelerated expansion of the Universe\cite{SupernovaSearchTeam:1998,SupernovaCosmologyProject:1999}. `CDM' stands for cold dark matter, which accounts for 80\% of the matter density in the Universe\cite{peebles:2020}. The existence of dark matter and dark energy, as well as their mysterious and elusive nature, are well known to the public.

Yet, the most extraordinary ingredient of the model is perhaps not as well known. This ingredient, which is not contained in the name of $\Lambda$CDM, is the idea that the origin of all structures in the Universe, such as galaxies, stars, planets and life, was the quantum-mechanical vacuum fluctuation generated in the early Universe\cite{mukhanov/chibisov:1981,hawking:1982,starobinsky:1982,guth/pi:1982,bardeen/turner/steinhardt:1983}.
The observed properties of cosmic structures, most notably those of the afterglow of the fireball Universe called the cosmic microwave background (CMB), agree with this idea\cite{komatsu/etal:2014,Planck2018X}.

What is the physical nature of dark matter and dark energy? What generated the initial quantum vacuum fluctuation and how? {\it Polarised} light of the CMB may help answer these questions. Their importance is widely recognised, and there are beautiful measurements of CMB polarisation from two space missions,
the National Aeronautics and Space Administration (NASA) \textit{Wilkinson Microwave Anisotropy Probe} (\textit{WMAP})\cite{wmap:2013a} and the European Space Agency (ESA) \textit{Planck}\cite{Planck2018I}, as well as from a host of ground-based\cite{adachi/etal:2020,adachi/etal:2020b,aiola/etal:2020,sayre/etal:2020,dutcher/:2021,BICEP:2021} and balloon-borne\cite{SPIDER:2022} experiments. The rate at which the sensitivity of CMB experiments has improved over the last 2 decades is staggering: the noise level has dropped by 3 orders of magnitude, nearly exponentially with time. If it were not for the CMB research, such a dramatic improvement and innovation in microwave sensor technology would not have been made.

There will be ever more powerful CMB polarisation experiments in the coming decade, including ground-based observatories such as
Simons Observatory\cite{SimonsObservatory:2019}, South Pole Observatory\cite{SPO:2020} and CMB Stage-4\cite{CMB-S4:2019}, and a space mission \textit{LiteBIRD}\cite{LiteBIRD:2022} led by the Japan Aerospace Exploration Agency (JAXA). These experiments will reduce the noise level by another order of magnitude. We are now at the stage where not only the statistical uncertainty but also the systematic uncertainty (including both instrumental and astrophysical ones) must be controlled with unprecedented precision. What kind of systematics should we characterise better and how well depends on what kind of new physics we wish to discover from such observations.

In this article, we focus on two new developments in the quest for new physics. First, CMB polarisation is sensitive to physics violating parity symmetry under inversion of spatial coordinates\cite{lue/wang/kamionkowski:1999}. We discuss a tantalising hint for such a signal, called `cosmic birefringence'\cite{carroll/field/jackiw:1990,carroll/field:1991,harari/sikivie:1992}, in the polarisation data obtained by \textit{Planck}\cite{minami/komatsu:2020b,NPIPE:2022,eskilt:prep}. The statistical significance is currently about 3$\sigma$. If confirmed with higher significance in future, it would have profound implications for the fundamental physics behind dark energy\cite{carroll:1998,panda/sumitomo/trivedi:2011} and dark matter\cite{finelli/galaverni:2009,fedderke/graham/rajendran:2019}, as well as for theory of quantum gravity\cite{myers/pospelov:2003,arvanitaki/etal:2010}.

Second, the quantum-mechanical vacuum fluctuation generated in the early Universe could produce a stochastic background of primordial gravitational waves (GW)\cite{grishchuk:1975,starobinsky:1979}, and CMB polarisation is sensitive to them\cite{seljak/zaldarriaga:1997,kamionkowski/etal:1997b}. We discuss recent new developments in the study of GW sourced by matter fields in the early Universe. Statistical properties of such a sourced GW are dramatically different from those of the quantum vacuum fluctuation in spacetime (i.e., tensor metric perturbation) that has usually been considered in the CMB research\cite{kamionkowski/kovetz:2016}.

These two developments may as well require new designs and calibration strategy for future CMB experiments, and influence the way polarisation data are analysed. We conclude this article by giving an outlook in this regard.

\section*{Polarisation of the CMB}
Space is filled with nearly uniform sea of CMB photons from the fireball Universe\cite{peebles/page/partridge:2009}.
When the temperature was greater than 3,000~K in the past, all atoms were fully ionised and photons were scattered by electrons efficiently: the plasma was opaque to photons. When the temperature fell below 3,000~K, protons, helium nuclei and electrons were combined to form neutral atoms, which did not scatter photons very much. This marks the moment of `last scattering', after which most of the photons propagated freely 13--14 billion years to reach us today.

As the photons propagated through expanding space, their wavelength was stretched and the photons lost energy continuously to expansion. The energy spectrum of photons remained a thermal Planck spectrum (which was established by interaction of matter and radiation in the fireball phase\cite{sunyaev/zeldovich:1970,danese/dezotti:1982}) with temperature dropping as space expanded. The present-day temperature averaged over the full sky is $T_0=2.725$~K\cite{fixsen:2009}, and the spectrum has a peak in microwave bands, hence the name `cosmic microwave background'.

The initial fluctuations generated in the early Universe left imprints in the distribution of CMB intensity observed in different directions on the sky. As the spectrum of CMB intensity is a Planck spectrum, the results of CMB experiments are reported in units of thermodynamic temperature, rather than in units of intensity. The temperature fluctuation observed in a direction $\hat n$ is defined by removing the sky-averaged temperature, $\Delta T(\hat n)=T(\hat n)-T_0$. A typical magnitude of the fluctuation is $\Delta T/T_0$ of order $10^{-5}$. By measuring the distribution of CMB photons arriving from various directions over the sky, we can map $\Delta T$ at the `surface of last scattering' far, far away from us.

The CMB photons are linearly polarised at the surface of last scattering\cite{kosowsky:1996}. Linear polarisation is generated when \textit{anisotropic} incident light is scattered/reflected. The sunlight reflected upon the windshield of a car is polarised because the sunlight arrives from the sky. A rainbow is polarised when the sunlight is scattered by water droplets in the atmosphere into a particular direction. How about the Universe? The Universe is isotropic with no preferred direction, but the angular distribution of intensity of light coming to an electron can be \textit{locally} anisotropic. Linear polarisation is generated when this anisotropic incident light is last scattered by electrons\cite{hu/white:1997}.

\begin{figure}[ht]
\centering
\includegraphics[width=\linewidth]{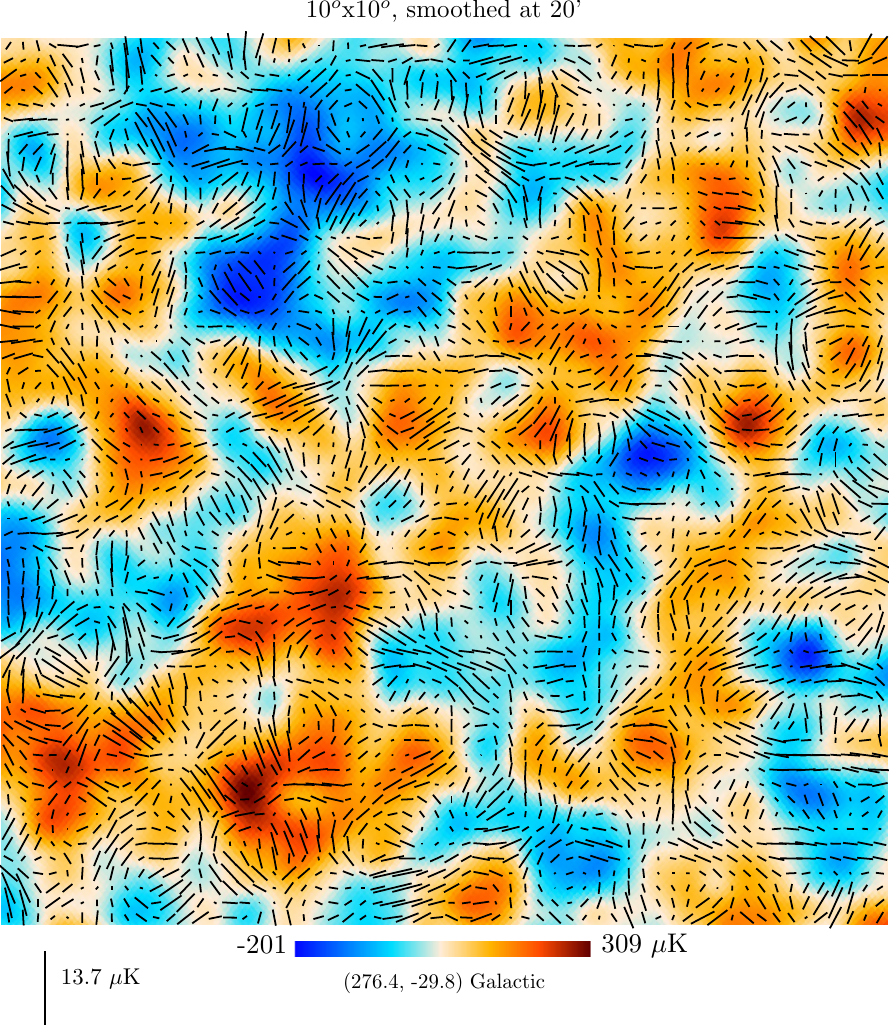}
\caption{{\bf Observed CMB polarisation.}
This map shows temperature fluctuations $\Delta T$ (colour) and polarisation (lines) on a $10^\circ \times 10^\circ$ patch of sky, extracted from the full-sky map of the \textit{Planck} mission. Lengths and orientations of the lines indicate polarisation strengths and directions, respectively. The figure is reproduced from Ref.\cite{Planck2018I}.}
\label{fig:planck}
\end{figure}

In Fig.~\ref{fig:planck}, we show a high signal-to-noise map of CMB polarisation observed by the \textit{Planck} mission\cite{Planck2018I} overlaid on a map of $\Delta T$. In this article, we describe how such a polarisation map can tell us about new physics.

\subsection*{$E$- and $B$-mode polarisation and parity}
To probe violation of parity symmetry, we decompose the observed pattern of CMB polarisation into eigenstates of parity, called $E$ and $B$ modes\cite{zaldarriaga/seljak:1997,kamionkowski/etal:1997}, which transform differently under inversion of spatial coordinates. Let us first consider a small patch of sky (Fig.~\ref{fig:planck}) around a given line of sight.

We use Stokes parameters to characterise the polarisation field. In right-handed coordinates with the $z$ axis taken in the propagation direction of photons, Stokes parameters for linear polarisation are $Q=|E_x|^2-|E_y|^2=2\Re(E_+^*E_-)$ and $U=2\Re(E_x^*E_y)=2\Im(E_+^*E_-)$, where $E_x$ and $E_y$ are components of an electric field $\bm{E}$ in Cartesian coordinates and
$E_{\pm}= (E_x\mp iE_y)/\sqrt{2}$ are those of $\pm$ helicity states. Right- and left-handed circular polarisation states correspond to $+$ and $-$, respectively.

With an amplitude ($P$) and a position angle (PA) of linear polarisation defined as $Q\pm iU=Pe^{\pm 2i\psi}$, $\psi>0$ is a counter-clockwise rotation of the plane of polarisation on the sky. This is the definition of PA adopted by the International Astronomical Union (IAU)\cite{hamaker/bregman:1996}. On the other hand, the CMB community often uses right-handed coordinates with the $z$ axis taken in the direction of observer's lines of sight. This results in the opposite sign convention for $U$ and PA\cite{angle:2017}.

We adopt the CMB convention for the rest of this article and use the notation $\beta=-\psi$ for PA, as this is what has been reported in the literature. In this notation $\beta>0$ is a \textit{clockwise} rotation on the sky, and
$Q\pm iU=Pe^{\pm 2i\beta}=Pe^{\mp 2i\psi}=Q_{\mathrm{IAU}}\mp iU_{\mathrm{IAU}}$. The Stokes parameter for circular polarisation, $V=2\Im(E_x^*E_y)=|E_+|^2-|E_-|^2$, is not affected by this difference.

Without loss of generality, we choose the centre of the sky patch to be at the pole of spherical coordinates ($\theta=0$).
We define $E$ and $B$ modes by writing Fourier transform of Stokes parameters observed at a 2-dimensional position vector, $\bm{\theta}=(\theta\cos\phi,\theta\sin\phi)$, as
\begin{equation}
\label{eq:EB}
Q(\bm{\theta})\pm iU(\bm{\theta})=\int \frac{d^2\bm{\ell}}{(2\pi)^2}\left(E_{\bm{\ell}}\pm iB_{\bm{\ell}}\right)e^{\pm 2i\phi_{\bm{\ell}}+i\bm{\ell}\cdot\bm{\theta}}\,,
\end{equation}
where $\bm{\ell}=(\ell\cos\phi_{\bm{\ell}},\ell\sin\phi_{\bm{\ell}})$. Here, $E_{\bm{\ell}}$ and $B_{\bm{\ell}}$ are Fourier coefficients of $E$ and $B$ modes, respectively. For a single plane wave, $E_{\bm{\ell}}$ produces polarisation directions that are parallel or perpendicular to $\bm{\ell}$, and $B_{\bm{\ell}}$ produces those tilted by $\pm 45^\circ$. By construction $E_{\bm{\ell}}$ has even parity, whereas $B_{\bm{\ell}}$ has odd parity.

We acknowledge the abuse of notation: $E$- and $B$-mode polarisation coefficients, $E_{\bm{\ell}}$ and $B_{\bm{\ell}}$, must not be confused with electric and magnetic field vectors, $\bm{E}$ and $\bm{B}$! In fact, $\bm{E}$ is a vector with odd parity, and $\bm{B}$ is a pseudovector with even parity.

Fourier transform given in equation~(\ref{eq:EB}) is valid only on a small patch of sky. To deal with full-sky data on a sphere, we use spherical harmonics. We expand Stokes parameters observed in a direction $\hat n=(\sin\theta\cos\phi,\sin\theta\sin\phi,\cos\theta)$ on the sky as
\begin{equation}
\label{eq:EBfullsky}
    Q(\hat{n})\pm iU(\hat n)=-\sum_{\ell=2}^\infty\sum_{m=-\ell}^{\ell}\left(E_{\ell m}\pm iB_{\ell m}\right) {}_{\pm 2}Y_\ell^m(\hat n)\,,
\end{equation}
where ${}_{\pm 2}Y_\ell^m(\hat n)$ is the spin-2 spherical harmonics\cite{zaldarriaga/seljak:1997,kamionkowski/etal:1997}, and $E_{\ell m}$ and $B_{\ell m}$ are spherical harmonics coefficients of $E$ and $B$ modes, respectively. This is the full-sky generalisation of equation~(\ref{eq:EB}) and is used for analysing polarisation data of all experiments. Under $\hat n\to -\hat n$, the coefficients transform as $E_{\ell m}\to (-1)^\ell E_{\ell m}$ and $B_{\ell m}\to (-1)^{\ell+1} B_{\ell m}$; thus, they have the opposite parity, as promised. In the limit of small angles (large $\ell$), $\ell$ is equal to $|\bm{\ell}|$, and $m$ describes how the Fourier coefficients depend on $\phi_{\bm{\ell}}$.

\begin{figure*}[ht]
\centering
\includegraphics[width=\linewidth]{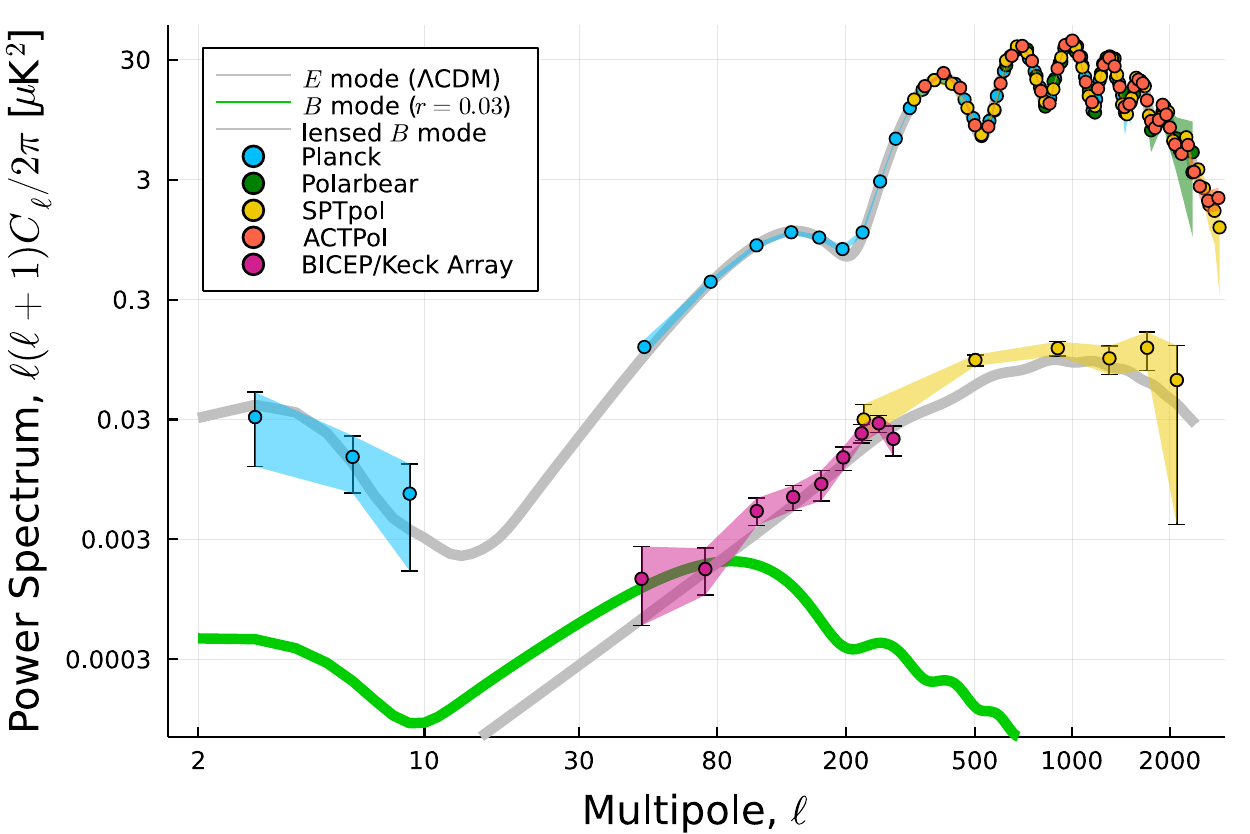}
\caption{{\bf Observed CMB polarisation power spectra.} We show $\ell(\ell+1)C_\ell^{EE}/2\pi$ and $\ell(\ell+1)C_\ell^{BB}/2\pi$ in units of $\mu\mathrm{K}^2$ from the current generation of CMB experiments: \textit{Planck}\cite{Planck2018I}, Polarbear\cite{adachi/etal:2020b}, South Pole Telescope (SPTpol)\cite{sayre/etal:2020,dutcher/:2021}, Atacama Cosmology Telescope (ACTPol)\cite{choi/etal:2020}, and BICEP/Keck Array\cite{BICEP:2021}. The data points are averaged over multipoles in bins centered at $\ell$ shown by the filled circles. The error bars and bands indicate the 68\% confidence level (C.L.) intervals. The best-fitting $\Lambda$CDM models\cite{Planck2018VI} for the $E$-mode and lensed $B$-mode power spectra are also shown, together with the $B$-mode power spectrum of the primordial GW for the tensor-to-scalar parameter of $r=0.03$.}
\label{fig:cl}
\end{figure*}

A useful quantity for describing stochastic variables such as $E_{\ell m}$ and $B_{\ell m}$ is the angular power spectrum $C_\ell$, which is the squared amplitude of spherical harmonics coefficients. Assuming that the Universe is statistically isotropic with no preferred direction, we average the squared coefficients over $m$ to obtain
\begin{align}
    C_\ell^{EE}&=\frac1{2\ell+1}\sum_{m=-\ell}^\ell|E_{\ell m}|^2\,,\\
    C_\ell^{BB}&=\frac1{2\ell+1}\sum_{m=-\ell}^\ell|B_{\ell m}|^2\,.
\end{align}
Both of these are parity even. See Fig.~\ref{fig:cl} for the current measurements of $C_\ell^{EE}$ and $C_\ell^{BB}$. Though not shown, the parity-even cross-correlation of temperature and polarisation, $C_\ell^{TE}$, has also been measured precisely.

The parity-odd power spectra (not shown in Fig.~\ref{fig:cl}) are given by
\begin{align}
C_\ell^{EB}&=\frac1{2\ell+1}\sum_{m=-\ell}^\ell \Re\left(E_{\ell m}B^*_{\ell m}\right)\,,\\
C_\ell^{TB}&=\frac1{2\ell+1}\sum_{m=-\ell}^\ell \Re\left(T_{\ell m}B^*_{\ell m}\right)\,,
\end{align}
and can be used to probe new physics that violates parity symmetry\cite{lue/wang/kamionkowski:1999}. While $C_\ell^{TB}$ used to be the most sensitive probe of parity violation in the \textit{WMAP} era\cite{WMAP:2009,WMAP:2011}, $C_\ell^{EB}$ has become the most sensitive one in the current era of CMB experiments with low polarisation noise.

The $C_\ell^{EE}$ data are dominated by sound waves excited by density fluctuations in the fireball Universe. Photons and electrons were tightly coupled via Thomson scattering, and electrons, protons and helium nuclei were also tightly coupled via Coulomb scattering; thus, the cosmic plasma behaved as if it were a single fluid, i.e., a `cosmic hot soup'\cite{peebles/yu:1970,sunyaev/zeldovich:1970b}. Density fluctuations excited sound waves in this fluid, which have been observed clearly as peaks and troughs in $C_\ell^{TT}$\cite{miller/etal:1999,boomerang:2000,hanany/etal:2000} and $C_\ell^{TE}$\cite{WMAP:2003} as well as in $C_\ell^{EE}$ shown here. The early Universe was indeed filled with sound waves propagating in a hot soup.

Non-linear effects such as the gravitational lensing effect of CMB by the intervening matter distribution in the Universe mix $E$ and $B$ modes at different multipoles\cite{zaldarriaga/seljak:1998} and produce non-zero $C_\ell^{BB}$. No $C_\ell^{EB}$ is generated by this process unless parity symmetry is violated by other new physics, as we describe in the next section. This lensing-induced $C_\ell^{BB}$ has been measured as shown in Fig.~\ref{fig:cl}.

$B$ modes can also be generated by primordial GW\cite{seljak/zaldarriaga:1997,kamionkowski/etal:1997b}, which we discuss in the later sections. Using $C_\ell^{EE}$ and $C_\ell^{BB}$ shown in Fig.~\ref{fig:cl}, we find that the map shown in Fig.~\ref{fig:planck} is consistent with no $B$ modes from GW\cite{tristram/etal:2021}.
The lensing $B$ mode may be a nuisance in the search of primordial GW, but it is a treasure when investigating the mass distribution (including dark matter) in the Universe\cite{omori/etal:2017,Planck2018VIII,POLARBEAR:2020,darwish/etal:2020}.

\section*{New Physics I: Cosmic birefringence}
There exists a distinct possibility that a parity-violating pseudoscalar field, $\chi$, is responsible for dark matter and dark energy\cite{marsh:2016,ferreira:2021}. The concept of a pseudoscalar field, which changes sign under inversion of spatial coordinates, is familiar in particle physics. In SM, pion is a pseudoscalar\cite{chinowsky/steinberger:1954}. In beyond SM, the strong \textit{CP} problem of quantum chromodynamics (QCD) can be solved by introducing a yet-to-be-discovered pseudoscalar `axion' field\cite{peccei/quinn:1977,weinberg:1978,wilczek:1978}.

In this article, $\chi$ is some new pseudoscalar which can be fundamental (`axionlike' field) or composite. Like pion and axion, $\chi$ couples to an electromagnetic (EM) field in a parity-violating manner. If $\chi$ is dark matter or dark energy, it affects polarisation of CMB as photons travel through space filled with $\chi$ for more than 13 billion years.

\subsection*{Light propagation in the expanding Universe}
To set the notation, we first review the basics of propagation of a free EM field in expanding space. The material covered in this section will also be used in the later sections on GW. The speed of light is $c\equiv 1$ throughout this article.

We choose coordinates in which a distance between two events in homogeneous and isotropic spacetime is given by $ds^2=\sum_{\mu\nu}g_{\mu\nu}dx^\mu dx^\nu=a^2(\eta)(-d\eta^2+d\bm{x}^2)$, with the metric tensor being $g_{\mu\nu}=a^2(\eta)\mathrm{diag}(-1,\bm{1})$.
Here, $\bm{x}$ denotes comoving coordinates, whereas $\eta=x^0$ is the conformal time which is related to the physical time $t$ as $\eta=\int dt/a(t)$, with $a(t)$ being the scale factor of expanding space.

The action of a free EM field is $I_A=\int d^4x\sqrt{-g}{\cal L}_A$ with a Lagrangian density ${\cal L}_A=-\sum_{\mu\nu}F_{\mu\nu}F^{\mu\nu}/4$, where $F_{\mu\nu}\equiv \partial A_\nu/\partial x^\mu-\partial A_\mu/\partial x^\nu$ is the antisymmetric electromagnetic tensor and $g$ is the determinant of $g_{\mu\nu}$.
With gauge conditions $A_0=0$ and $\nabla\cdot\bm{A}=0$, the equation of motion for $\bm{A}$ is given by $\bm{A}''-\nabla^2\bm{A}=0$, where $\nabla\equiv\partial/\partial\bm{x}$ and the prime denotes $\partial/\partial\eta$.

Those who are not familiar with electromagnetism in cosmology may be surprised to see that the equation of motion for $\bm{A}$ takes the same form as in flat Minkowski space. The fundamental reason is that a massless free vector field is \textit{conformally coupled} to gravitation, which implies that, when written in suitable coordinates ($\eta$, $\bm{x}$), it does not `feel' expansion of space and behaves as if it were in Minkowski space. Conformal transformation rescales the metric tensor as $g_{\mu\nu}\to \hat g_{\mu\nu}=\Omega^2 g_{\mu\nu}$. For example, if we choose $\Omega=a^{-1}$, we can `undo' expansion and the transformed metric is equal to the Minkowski metric, $\hat g_{\mu\nu}=\mathrm{diag}(-1,\bm{1})$. Transformation yields $\sqrt{-\hat g}=\Omega^4\sqrt{-g}$ and $\hat F^{\mu\nu}=\Omega^{-4}F^{\mu\nu}$; thus, $\sqrt{-g}{\cal L}_A$ remains invariant. We can calculate everything in Minkowski space and the result is valid for all conformally transformed metric tensors in suitable coordinates.

In Fourier space the equation of motion is $\bm{A}''+k^2\bm{A}=0$, where $k$ is a comoving wavenumber, which is related to a physical wavelength as $\lambda(t)=2\pi a(t)/k$ (i.e., the wavelength of photons gets redshifted in physical coordinates). The equation of motion for $\pm$ helicity states is $A''_\pm+k^2 A_\pm=0$, yielding the identical dispersion relation for both states.

The electric and magnetic fields are given by $F_{i0}=a^2E_i$ (i.e., $\bm{E}=-a^{-2}\bm{A}'$) and $F_{ij}=a^2\sum_{k}\epsilon^{ijk}B_k$ (i.e., $\bm{B}=a^{-2}\nabla\times\bm{A}$), respectively. Here, $\epsilon^{ijk}$ is a totally antisymmetric symbol with $\epsilon^{123}=1$. We then find ${\cal L}_A=(\bm{E}\cdot\bm{E}-\bm{B}\cdot\bm{B})/2$. In our notation, $\bm{E}\cdot\bm{E}=\sum_iE_i^2$.
The stress-energy tensor is $T_{ij}^A=\sum_{\mu\nu}g^{\mu\nu}F_{i\mu}F_{j\nu}+g_{ij}{\cal L}_A$, which reproduces the known result for EM pressure, $P_A=(\bm{E}\cdot\bm{E}+\bm{B}\cdot\bm{B})/6=\rho_A/3$, where $\rho_A$ is the energy density. Conformal invariance occurs generally for any stress-energy sources with a vanishing trace, $\sum_{\mu\nu}g^{\mu\nu}T_{\mu\nu}=0$, which is certainly the case for $T_{\mu\nu}^A$. For a general perfect fluid, a vanishing trace implies $P=\rho/3$.

\subsection*{Rotation of the plane of linear polarisation}
We now include $\chi$. A pseudoscalar can couple to EM via the so-called Chern-Simons (CS) term in the action\cite{ni:1977,turner/widrow:1988}
\begin{align}
\label{eq:CS}
    I_\mathrm{CS} &= \int d^4x \sqrt{-g}\left(-\frac{\alpha}{4f}\chi F\tilde F\right)\,,\\
    F\tilde F&\equiv \sum_{\mu\nu}F_{\mu\nu}\sum_{\mu'\nu'}\frac{\epsilon^{\mu\nu\mu'\nu'}}{2\sqrt{-g}}F_{\mu'\nu'}\,,
\end{align}
where $\alpha$ is a dimensionless coupling constant, $f$ is the so-called `decay constant' with dimension of energy, and $\epsilon^{\mu\nu\mu'\nu'}$ is a totally antisymmetric symbol with $\epsilon^{0123}=1$. The $F\tilde F$ term violates parity symmetry, as $F\tilde F=-4\bm{B}\cdot\bm{E}$ changes sign under inversion of spatial coordinates. Therefore, $\chi$ needs to be a pseudoscalar, such that the whole $\chi F\tilde F$ remains invariant.

In SM, a neutral pion $\pi_0$ decays into 2 photons via this coupling in the effective Lagrangian density $-2g_\pi\pi_0F\tilde F$ with $g_\pi=N_ce^2/(48\pi^2f_\pi)$\cite{weinberg:1996}; thus, $\alpha=N_ce^2/(6\pi^2)$ for $f=f_\pi$ where $N_c=3$ is the number of colours and $f_\pi\simeq 184$~MeV is the pion decay constant. For our purpose, $\alpha/f$ is a free parameter related to the physical nature of dark matter or dark energy, which can be constrained by CMB polarisation data.

For simplicity, we first discuss the effect of homogeneous $\chi=\chi(\eta)$. This is a good approximation for dark energy\cite{carroll:1998} but may not apply to other cases. We discuss spatial fluctuations later. The equation of motion for $\bm{A}$ is given by $\bm{A}''-\nabla^2\bm{A}-(\alpha\chi'/f)\nabla\times\bm{A}=0$, and that for $A_\pm$ is given by\cite{carroll/field/jackiw:1990,carroll/field:1991,harari/sikivie:1992}
\begin{equation}
\label{eq:vectorequation}
    A''_{\pm} +\left(k^2\mp \frac{k\alpha\chi'}{f}\right)A_{\pm} = 0\,.
\end{equation}
The effect of the CS term vanishes when $\chi$ is a constant, because $\sqrt{-g}F\tilde{F}$ is a total derivative and does not contribute to physics unless the coefficient multiplying it, $\chi$, depends on spacetime.

The equation of motion now yields a helicity-dependent dispersion relation. When the effective angular frequency, $\omega_\pm^2\equiv k^2\mp k\alpha\chi'/f$, varies slowly with time within one period, $|\omega_{\pm}'|/\omega_{\pm}^2\ll 1$, a WKB solution is $A_{\pm}\simeq
(2\omega_{\pm})^{-1/2}e^{-i\int d\eta\omega_{\pm}+i\delta_{\pm}}$, where $\delta_{\pm}$ is the initial phase of the $\pm$ states.
The phase velocity is given by
\begin{equation}
\label{eq:phasevelocity}
\frac{\omega_\pm}{k}\simeq 1\mp \frac{\alpha\chi'}{2kf}\,,
\end{equation}
when the second term is small. Indeed, the second term is absolutely tiny: it is at most of order the ratio of the photon wavelength and the size of the visible Universe, $(k\eta)^{-1}$. However, the impact on CMB accumulates over very long time (more than 13 billion years), which makes $\int d\eta(\omega_+-\omega_-)$ large enough to be observable. Therefore, we keep $\omega_\pm$ in the phase but set $\omega_{\pm}\simeq k$ in the amplitude of the WKB solution.

On the other hand, the second term in $\omega_\pm^2= k^2\mp k\alpha\chi'/f$ can exceed the first term during the period of cosmic inflation, resulting in $\omega_\pm^2<0$ depending on the sign of $\alpha$\cite{anber/sorbo:2010}. This signals instability, and yields rich phenomenology for primordial GW\cite{sorbo:2011,anber/sorbo:2012,barnaby/pajer/peloso:2012}. We discuss this in the later sections on GW.

 The group velocity is not modified at linear order of $\alpha$ but receives a very tiny positive contribution at $\alpha^2$, making $d\omega_{\pm}/dk\simeq 1+(\alpha\chi'/f)^2/(8k^2)>1$. Whether this has any significance is an open question\cite{mcdonald/ventura:2020}, although it has no practical consequence for our study.

The difference in the phase velocity leads to rotation of the plane of linear polarisation\cite{carroll/field/jackiw:1990,carroll/field:1991,harari/sikivie:1992}. Suppose that the initial light of CMB at the surface of last scattering had no circular polarisation, i.e., both helicity states had equal amplitudes of electric fields, $E_\pm\propto e^{-i\int d\eta\omega_{\pm}+i\delta_{\pm}}$. In the CMB convention defined earlier,
the Stokes parameters for linear polarisation are $Q\propto \cos[\int d\eta(\omega_+-\omega_-)-(\delta_+-\delta_-)]$ and $U\propto -\sin[\int d\eta(\omega_+-\omega_-)-(\delta_+-\delta_-)]$. The PA is given by
\begin{equation}
\label{eq:beta}
    \beta=\frac12\left(\delta_+-\delta_-\right)-\frac12\int d\eta\left(\omega_+-\omega_-\right)\,.
\end{equation}
The plane of linear polarisation is rotated relative to the initial PA at the surface of last scattering, $(\delta_+-\delta_-)/2$. This phenomenon is similar to birefringence in a crystal.

The Stokes parameter for circular polarisation vanishes, $V=0$, as we set $\omega_\pm\simeq k$ in the amplitude of the WKB solution. Keeping $\omega_{\pm}$ results in a tiny $V$\cite{finelli/galaverni:2009}. Non-zero $V$ occurs at the order $\alpha^2$ as well\cite{alexander/ochoa/kosowsky:2009}.

Without loss of generality, we set $\delta_+-\delta_-=0$ from now on. Using equation~(\ref{eq:phasevelocity}), we find $\beta=(\alpha/2f)\int_{\eta_{\mathrm{LS}}}^{\eta_0} d\eta\chi'$, where the subscripts `LS' and `0' denote the time of last scattering and the present-day time, respectively. Therefore, space filled with a time-dependent $\chi$, which could be dark matter or dark energy or both, behaves as if it were a birefringent material. For this reason, such an effect is called `cosmic birefringence'.

\begin{figure}[ht]
\centering
\includegraphics[width=\linewidth]{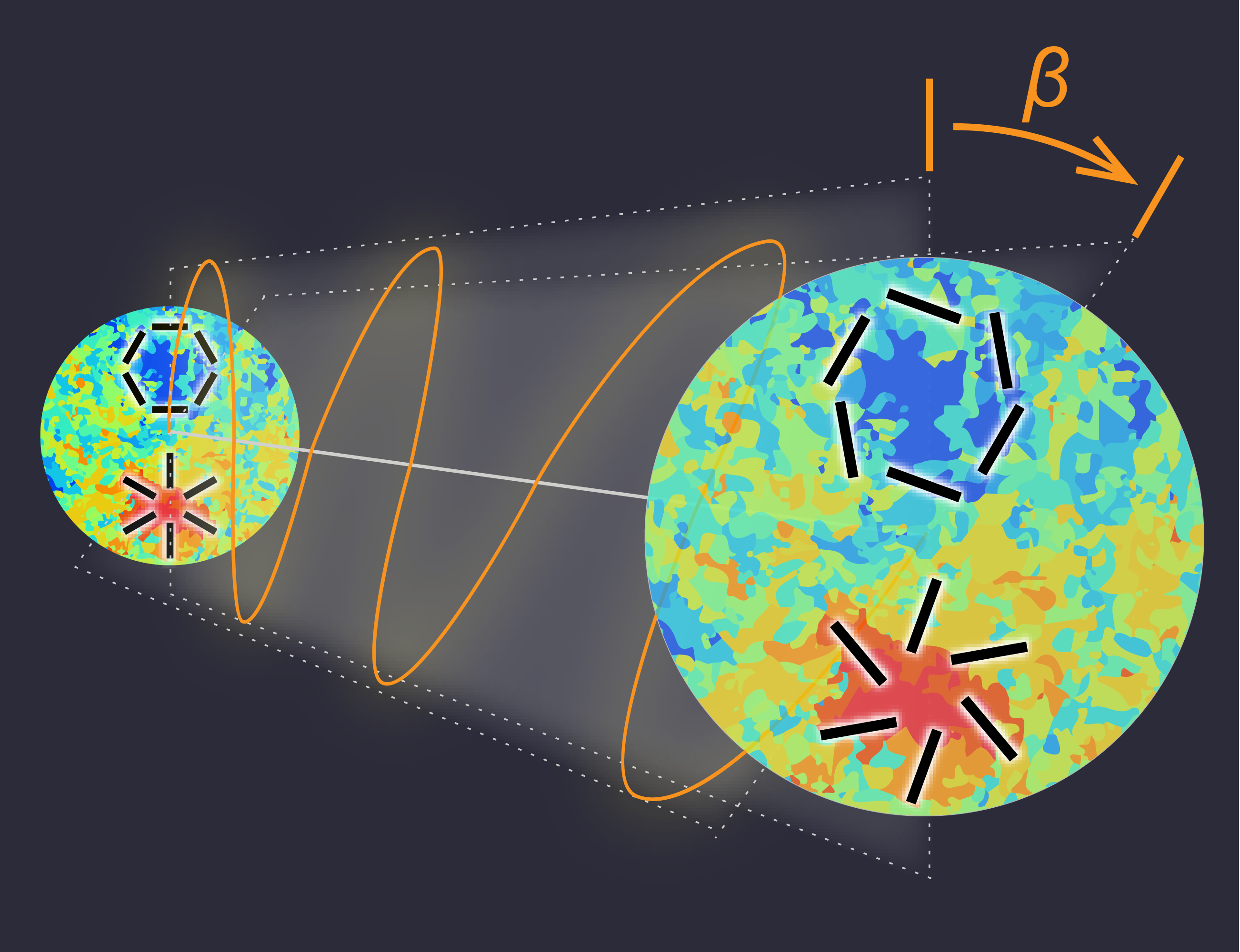}
\caption{{\bf Rotation of linear polarisation by cosmic birefringence.} Polarisation pattern at the surface of last scattering (black lines inside the left circle) changes when the plane of linear polarisation (orange line) rotates by an angle $\beta$ as CMB photons travel to reach us today (right circle). Credit: Y. Minami.}
\label{fig:minami}
\end{figure}

We sketch this effect in Fig.~\ref{fig:minami}: $\beta>0$ is a clockwise rotation on the sky. If we write ${\cal L}_\mathrm{CS}=\frac14g_a\chi F\tilde F$ for a Lagrangian density of the CS term\cite{harari/sikivie:1992} instead of that in equation~(\ref{eq:CS}), $\beta=-\frac12g_a\int_{\eta_{\mathrm{LS}}}^{\eta_0} d\eta\chi'$.

We can derive the same result in an alternative way\cite{harari/sikivie:1992}. Writing the Lagrangian density,
\begin{align}
\nonumber
    &{\cal L}_A+{\cal L}_{\mathrm{CS}}
    =\frac12\left(\bm{E}\cdot\bm{E}-\bm{B}\cdot\bm{B}\right)
    +\frac{\alpha}{f}\chi\left(\bm{B}\cdot\bm{E}\right)\\
    &=\frac12\left[\left(\bm{E}+\frac{\alpha}{2f}\chi\bm{B}\right)^2-\left(\bm{B}-\frac{\alpha}{2f}\chi\bm{E}\right)^2\right]+{\cal O}(\alpha^2)\,,
\end{align}
we find that the linear combinations $\bm{E}+(\alpha/2f)\chi\bm{B}$
and $\bm{B}-(\alpha/2f)\chi\bm{E}$ satisfy free wave equations and oscillate along a constant direction. The plane of linear polarisation therefore rotates clockwise on the sky by an angle $\beta=(\alpha/2f)\left[\chi(\eta_0)-\chi(\eta_{\mathrm{LS}})\right]$.

We now include spatial fluctuations, $\chi(\eta)\to\chi(\eta,\bm{x})$. Conveniently, all we need is to replace
$k\chi'\to k\chi'+\bm{k}\cdot\nabla\chi$ in Eq.~(\ref{eq:vectorequation}), i.e., the total derivative of $\chi$ along the photon trajectory\cite{harari/sikivie:1992,li/zhang:2008,pospelov/ritz/skordis:2009}.
The birefringence angle is given by
\begin{equation}
\label{eq:betahat}
    \beta(\hat n)=\frac{\alpha}{2f}\left[\chi(\eta_0)-\chi(\eta_{\mathrm{LS}},r_{\mathrm{LS}}\hat n)\right]\,,
\end{equation}
where 
$r_{\mathrm{LS}}\equiv \eta_0-\eta_{\mathrm{LS}}$ is the comoving distance from the observer to the surface of last scattering and $\hat n$ is the direction of observer's lines of sight.

Fluctuations in $\beta(\hat n)$ are easier to measure experimentally than the isotropic $\beta$, as they do not require knowledge of the absolute PA of polarisation-sensitive detectors on the focal plane with respect to the sky\cite{kamionkowski:2009}. Currently there is no evidence for fluctuations in $\beta(\hat n)$\cite{contreras/boubel/scott:2017,bianchini/etal:2020,namikawa/etal:2020,gruppuso/etal:2020}; thus, we focus on the isotropic $\beta$ for the rest of this article.

The phenomenon of cosmic birefringence can occur more generally. Integrating equation~(\ref{eq:CS}) by parts, we obtain
\begin{equation}
	I_{\mathrm{CS}}=\frac{\alpha}{2f}\int d^4x\sqrt{-g}
    \sum_{\mu\nu}p_\mu A_\nu \tilde F^{\mu\nu}\,,
\end{equation}
where $p_\mu=\partial\chi/\partial x^\mu$ for the pseudoscalar example. We now promote $p_\mu$ to a generic 4-vector. Since $p_\mu$ picks up a preferred direction in spacetime, it breaks Lorentz and \textit{CPT} symmetry; thus, cosmic birefringence is a probe of new physics that breaks Lorentz and \textit{CPT} symmetry\cite{carroll/field/jackiw:1990,colladay/kostelecky:1998,kostelecky/mewes:2002,kostelecky/mewes:2007}. For the homogeneous pseudoscalar example, cosmological evolution of $\chi$ picks up a preferred  direction $p_0=\chi'$. Lorentz breaking terms in the effective Lagrangian can generate non-zero circular polarisation\cite{alexander/ochoa/kosowsky:2009,zarei/etal:2010}.

So far we have considered $\beta$ that is independent of the photon energy $k$.
Lorentz and \textit{CPT} breaking terms in the effective Lagrangian can give $\beta(k)\propto k^n$ with $n=1$\cite{shore:2005} and $2$\cite{gambini/pullin:1999,myers/pospelov:2003}. Such signatures may be a sign of new physics at the Planck energy scale, e.g., quantum gravity.
In the absence of Lorentz and \textit{CPT} breaking terms, astrophysical effects such as Faraday rotation due to the intergalactic magnetic field can produce a $k$-dependent rotation of the plane of linear polarisation with $n=-2$\cite{kosowsky/loeb:1996}.
If $\chi$ is dark matter with non-zero magnetic moment\cite{pospelov/terveldhuis:2000,sigurdson/etal:2004}, a Gyromagnetic Faraday effect can also produce $\beta$ in the presence of an external magnetic field\cite{gardner:2008}. Unlike the usual Faraday effect with $n=-2$, $\beta$ from a Gyromagnetic effect is independent of $k$.

See Refs.\cite{kahniashvili/durrer/maravin:2008,gubitosi/etal:2009,gubitosi/paci:2013} for early constraints on $\beta(k)$. Later we discuss the latest constraint on $n$ from the \textit{Planck} data\cite{eskilt:prep}.

\subsection*{Effects on $E$ and $B$ modes}
When PA shifts uniformly over the sky by $\beta$, we have $Q^{\mathrm{o}}\pm iU^{\mathrm{o}}=(Q\pm iU)e^{\pm 2i\beta}$, where the superscript `o' denotes the observed value and $Q\pm iU$ on the right hand side is the intrinsic one at the surface of last scattering. This yields
$E_{\ell m}^\mathrm{o}\pm iB_{\ell m}^\mathrm{o}=(E_{\ell m}\pm iB_{\ell m})e^{\pm 2i\beta}$, or
\begin{align}
\label{eq:EBrelation1}
E_{\ell m}^\mathrm{o}&= E_{\ell m}\cos(2\beta)-B_{\ell m}\sin(2\beta)\,,\\
\label{eq:EBrelation2}
B_{\ell m}^\mathrm{o}&= E_{\ell m}\sin(2\beta)+B_{\ell m}\cos(2\beta)\,.
\end{align}
This simple result assumes an instantaneous decoupling of CMB photons at $\eta_{\mathrm{LS}}$. In reality, the last scattering of photons occurred over a finite duration characterised by the so-called `visibility function', ${\cal V}(\eta)$, which has a peak at $\eta_\mathrm{LS}$ with some width. We need to integrate the Boltzmann equation for photons to take this effect into account.

For simplicity, let us consider only the density fluctuation and ignore GW. We expand $Q\pm iU$ in Fourier space with the wavenumber $\bm{q}$ taken in the $z$ axis.
Defining $\mu=\cos\theta$ for the propagation direction of photons in spherical coordinates,
we write the Boltzmann equation for
the Fourier coefficients,
${}_{\pm 2}\Delta_{P}(\eta,q,\mu)$, as\cite{liu/lee/ng:2006,finelli/galaverni:2009,gubitosi/martinelli/pagano:2014}
\begin{align}
\nonumber
{}_{\pm 2}\Delta_{P}'+iq\mu {}_{\pm 2}\Delta_{P}
=&-\tau'{}_{\pm 2}\Delta_{P}
+\sqrt{20\pi}{}_{\pm 2}Y_2^0(\mu)\frac{\sqrt{6}}{10}\tau'\Pi\\
&\pm 2i\beta'{}_{\pm 2}\Delta_{P}\,,
\label{eq:boltzmann}
\end{align}
where $\beta'=-(\omega_+-\omega_-)/2$ (see equation~(\ref{eq:beta})),
$\tau'\equiv a\sigma_{\mathrm{T}}n_e$,
$\sigma_{\mathrm{T}}$ is the Thomson scattering cross section, $n_e$ the electron number density, and $\Pi$ the polarisation source function\cite{zaldarriaga/seljak:1997}.

Expanding ${}_{\pm 2}\Delta_{P}$ using spin-2 spherical harmonics,
\begin{equation}
    {}_{\pm 2}\Delta_{P}(\eta,q,\mu)=
    \sum_\ell i^{-\ell}\sqrt{4\pi(2\ell+1)}{}_{\pm 2}\Delta_{P,\ell}(\eta,q){}_{\pm 2}Y_\ell^0(\mu)\,,
\end{equation}
and defining the coefficients of $E$ and $B$ modes in the same way as in equation~(\ref{eq:EBfullsky}),
${}_{\pm 2}\Delta_{P,\ell}=-(\Delta_{E,\ell}\pm i
\Delta_{B,\ell})$, we find solutions to equation~(\ref{eq:boltzmann}) as
\begin{align}
\label{eq:boltzmannsolutionE}
\Delta_{E,\ell}(\eta_0,q)
&=\frac34\sqrt{\frac{(\ell+2)!}{(\ell-2)!}}
\int_0^{\eta_0}d\eta{\cal V}\Pi\frac{j_\ell(x)}{x^2}\cos\left[2\beta(\eta)\right]\,,\\
\label{eq:boltzmannsolutionB}
\Delta_{B,\ell}(\eta_0,q)
&=\frac34\sqrt{\frac{(\ell+2)!}{(\ell-2)!}}
\int_0^{\eta_0}d\eta{\cal V}\Pi\frac{j_\ell(x)}{x^2}\sin\left[2\beta(\eta)\right]\,,
\end{align}
where $\beta(\eta)=\int_{\eta}^{\eta_0}d\eta\beta'$, ${\cal V}(\eta)\equiv a\sigma_\mathrm{T}n_e e^{-\tau(\eta)}$ is the visibility function with $\tau(\eta)=\int_{\eta}^{\eta_0}d\eta\tau'$, and $j_\ell(x)$ the spherical Bessel function with $x= q(\eta_0-\eta)$.

For a slowly varying $\beta(\eta)$ and
a very sharply peaked ${\cal V}(\eta)$, these solutions agree with equations~(\ref{eq:EBrelation1}) and (\ref{eq:EBrelation2}) without $B_{\ell m}$ on the right hand side, as we ignored  primordial $B$ modes from GW at the surface of last scattering. On the other hand, for a rapidly oscillating $\beta(\eta)$ within a finite width of ${\cal V}(\eta)$, the effect is `washed out' and a proper treatment is required. This happens when $\chi$ is dark matter and starts oscillating before decoupling\cite{finelli/galaverni:2009,fedderke/graham/rajendran:2019}. See Refs.\cite{lee/liu/ng:2016,capparelli/caldwell/melchiorri:2020} for treatment of spatially fluctuating $\beta$ in the Boltzmann equation.

The Boltzmann equation is also required when we include the effect of reionisation of hydrogen atoms in a late-time Universe\cite{zaldarriaga:1997}. The so-called `reionisation bump' at $\ell\lesssim 10$ (see Fig.~\ref{fig:cl}) was generated at a redshift of $z\simeq 10$, and experienced less cosmic birefringence because of a shorter path length of photons\cite{liu/lee/ng:2006,WMAP:2009}. We can use this effect to perform `tomography': the difference between cosmic birefringence inferred from $\ell\lesssim 10$ and that from higher $\ell$ tells us the evolution of $\chi$ between $z\simeq 10$ and  $z_\mathrm{LS}\simeq 1090$\cite{sherwin/namikawa:prep}. As the current analysis of the \textit{Planck} data\cite{minami/komatsu:2020b,NPIPE:2022,eskilt:prep} is based only on $\ell\geq 51$, we use equations~(\ref{eq:EBrelation1}) and (\ref{eq:EBrelation2}) for the rest of this article.

\subsection*{Measuring $\beta$ from the $EB$ correlation}
Cosmic birefringence yields non-zero observed $C_\ell^{EB,{\mathrm{o}}}$ even if there were no $C_\ell^{EB}$ at the surface of last scattering\cite{lue/wang/kamionkowski:1999,feng/etal:2005,liu/lee/ng:2006}. Using equations~(\ref{eq:EBrelation1}) and (\ref{eq:EBrelation2}), we find
\begin{align}
C_\ell^{EE,{\mathrm{o}}} &=
\cos^2(2\beta)C_\ell^{EE}+\sin^2(2\beta)C_\ell^{BB}-C_\ell^{EB}\sin(4\beta)\,,\\
C_\ell^{BB,{\mathrm{o}}} &=
\sin^2(2\beta)C_\ell^{EE}+\cos^2(2\beta)C_\ell^{BB}+C_\ell^{EB}\sin(4\beta)\,,\\
C_\ell^{EB,{\mathrm{o}}}&=\frac12\sin(4\beta)(C_\ell^{EE}-C_\ell^{BB})
+C_\ell^{EB}\cos(4\beta)\,.
\end{align}
Using the solutions of the Boltzmann equation~(\ref{eq:boltzmannsolutionE}) and (\ref{eq:boltzmannsolutionB}), we can also calculate
\begin{equation}
    C_\ell^{XY,\mathrm{o}}=4\pi\int d\ln q
{\cal P}_\mathrm{s}(q)\Delta_{X,\ell}(\eta_0,q)\Delta_{Y,\ell}(\eta_0,q)\,,
\end{equation}
when $C_\ell^{BB}=0=C_\ell^{EB}$ initially. Here,
${\cal P}_\mathrm{s}(q)$ is the power spectrum of the initial scalar curvature perturbation.

Rotation by $\beta$ does not create $EB$ when $C_\ell^{EE}=C_\ell^{BB}$. This makes sense because $\beta$ mixes $E$ and $B$ modes; we need an asymmetry between the amplitudes of $E$ and $B$ modes to produce a non-zero effect. For example, randomly oriented polarisation angles have $C_\ell^{EE}=C_\ell^{BB}$ and $C_\ell^{EB}=0$. Rotation by $\beta$ still keeps them random, hence $C_\ell^{EB,\mathrm{o}}=0$.

We can write $C_\ell^{EB,{\mathrm{o}}}$ in terms of $C_\ell^{EE,{\mathrm{o}}}$ and $C_\ell^{BB,{\mathrm{o}}}$\cite{zhao/etal:2015,gruppuso/etal:2016}
\begin{equation}
\label{eq:EBobs}
C_\ell^{EB,{\mathrm{o}}}=
    \frac12(C_\ell^{EE,{\mathrm{o}}}-C_\ell^{BB,{\mathrm{o}}})\tan(4\beta)+\frac{C_\ell^{EB}}{\cos(4\beta)}\,.
\end{equation}
As shown in Fig.~\ref{fig:cl}, there is a large asymmetry between $C_\ell^{EE,{\mathrm{o}}}$ and $C_\ell^{BB,{\mathrm{o}}}$ in our Universe, which makes the $EB$ correlation a sensitive probe of $\beta$.

The intrinsic $C_\ell^{EB}$ on the right hand side could be generated by parity-violating primordial GW from gauge fields, as discussed in the later sections of this article. We can distinguish between the effects of $\beta$ and the intrinsic $C_\ell^{EB}$ easily using the shape of $C_\ell^{EB}$\cite{gluscevic/kamionkowski:2010,thorne/etal:2018}.

However, the effect of $\beta$ is degenerate with an instrumental miscalibration of polarisation angles~\cite{wu/etal:2009,WMAP:2011}. Do we observe $C_\ell^{EB,{\mathrm{o}}}$ because of $\beta$, or because we do not know how polarisation-sensitive orientations of detectors on the focal plane of a telescope are related to the sky coordinates and how polarisation of the incoming light is rotated by optical components precisely enough? If we rotate the focal plane by a miscalibration angle $\alpha$, the observed PA shifts and generates non-zero  $C_\ell^{EB,{\mathrm{o}}}$ in the same way as $\beta$. As a result, in the absence of any other information, we can only determine the sum of the two angles, $\alpha+\beta$, which explains why the previous determinations of $\beta$ were spread over a wide range beyond the quoted statistical uncertainties (excluding  systematic uncertainties in the knowledge of $\alpha$)\cite{PlanckIntXLIX,adachi/etal:2020,bianchini/etal:2020,namikawa/etal:2020,choi/etal:2020}.

Is there information we can use to lift degeneracy between $\alpha$ and $\beta$? One approach is to calibrate $\alpha$ using astrophysical objects with known PA\cite{kaufman/keating/leon:2016,aumont/etal:2020,masi/etal:2021}. However, to determine PA of an object with sufficient precision, we must have calibrated detectors well enough for such a measurement in the first place. This has not been achieved much better than $0.5^\circ$\cite{takahashi/etal:2010,koopman:2018}.

In 2019, Minami et al. proposed a new way to solve this issue\cite{minami/etal:2019}. The magnitude of cosmic birefringence is proportional to the path length of photons when $\chi'$ varies slowly, which makes CMB photons an ideal target. Our sky also contains polarised microwave emission from intersteller gas within our own Galaxy, which is often called the Galactic `foreground' emission. As the path length of photons within our Galaxy is much smaller, we can safely ignore $\beta$ for the foreground emission. Thus, polarisation of the foreground (FG) is rotated only by $\alpha$, whereas that of CMB by $\alpha+\beta$
\begin{align}
\nonumber
E_{\ell m}^{{\mathrm{o}}}&=
E_{\ell m}^{{\mathrm{FG}}}\cos(2\alpha)-B_{\ell m}^{{\mathrm{FG}}}\sin(2\alpha)\\
&+E_{\ell m}^{{\rm CMB}}\cos(2\alpha+2\beta)-B_{\ell m}^{{\rm CMB}}\sin(2\alpha+2\beta)\,,\\
\nonumber
B_{\ell m}^{{\mathrm{o}}}&= E_{\ell m}^{{\mathrm{FG}}}\sin(2\alpha)+B_{\ell m}^{{\mathrm{FG}}}\cos(2\alpha)\\
&+E_{\ell m}^{{\rm CMB}}\sin(2\alpha+2\beta)+B_{\ell m}^{{\rm CMB}}\cos(2\alpha+2\beta)\,,
\end{align}
which gives
\begin{align}
\nonumber
     C_\ell^{EB,{\mathrm{o}}}
    &= \frac{\tan(4\alpha)}2\left( C_\ell^{EE,{\mathrm{o}}}- C_\ell^{BB,{\mathrm{o}}}\right)\\
    \nonumber
    &+ \frac{\sin(4\beta)}{2\cos(4\alpha)}\left( C_\ell^{EE,{\rm CMB}}- C_\ell^{BB,{\rm CMB}}\right)\\
    &+\frac1{\cos(4\alpha)} C_\ell^{EB,{\mathrm{FG}}} +\frac{\cos(4\beta)}{\cos(4\alpha)} C_\ell^{EB,{\rm CMB}}\,.
\end{align}
This formula allows us to determine $\alpha$ and $\beta$ simultaneously, as $C_\ell^{EE,{\rm CMB}}$ and $C_\ell^{BB,{\rm CMB}}$ are known precisely.
Note that the formula does not require any knowledge of $C_\ell^{EE,\mathrm{FG}}$ or $C_\ell^{BB,\mathrm{FG}}$, but $C_\ell^{EB,\mathrm{FG}}$ and $C_\ell^{EB,\mathrm{CMB}}$. We can ignore $C_\ell^{EB,\mathrm{CMB}}$ for sensitivity of the current experiments, but $C_\ell^{EB,\mathrm{FG}}$ needs to be taken into account when interpreting the measured value of $\beta$.

\subsection*{Results from the \textit{Planck} polarisation data}
Applying the methodology developed in Refs.\cite{minami/etal:2019,minami:2020,minami/komatsu:2020} to the \textit{Planck} high-frequency instrument (HFI) data at $\nu=100$, 143, 217 and 353~GHz released in 2018\cite{Planck2018III}, a weak signal of $\beta=0.35^\circ\pm 0.14^\circ$ was reported for nearly full-sky data\cite{minami/komatsu:2020b}. Throughout this section, we quote uncertainties at the 68\,\% C.L. Subsequently, more precise value, $\beta=0.30^\circ\pm 0.11^\circ$, was reported\cite{NPIPE:2022} using the latest reprocessing of the \textit{Planck} data called `\texttt{NPIPE}'\cite{PlanckIntLVII}.

These measurements were done assuming that $\beta$ was independent of $\nu$.
Eskilt\cite{eskilt:prep} expanded the analysis by constraining the photon frequency dependence of $\beta(\nu)$. To this end he included all of the polarised frequency channels of the \textit{Planck} data, including those of the low-frequency instrument (LFI) at 30, 44 and 70~GHz\cite{Planck2018II}.
Parametrising the frequency dependence as $\beta(\nu)\propto \nu^n$, he finds $n=-0.35^{+0.48}_{-0.47}$, which is consistent with a frequency-independent $\beta$. Assuming $n=0$, he finds $\beta=0.33^\circ\pm 0.10^\circ$. The statistical significance exceeds $3\sigma$.
While intriguing, it is  not yet sufficient to claim a convincing detection.

If confirmed as a genuine signal of cosmic birefringence with more statistical significance in future, these measurements would provide evidence for new physics beyond SM with profound implications for the fundamental physics behind dark matter and dark energy. First of all, simply knowing that the physics behind them violates parity symmetry is a breakthrough. Next, as $\chi$ must evolve in time,
it would rule out Einstein's cosmological constant, $\Lambda$, as the origin of dark energy. Such a discovery would have a far-reaching consequence for quantum gravity: As $\Lambda>0$ is difficult to realise in quantum gravity\cite{dvali/gomez:2016,dvali:2020}, a recent `Swampland' proposal\cite{obied/etal:prep,garg/krishnan:2019,ooguri/etal:2019} favours a dynamical scalar field as the origin of dark energy\cite{agrawal/etal:2018,dvali/gomez/zell:2019}.
If $\chi$ is dark matter, $\beta$ may change during the course of observations as $\chi(\eta_0)$ in equation~(\ref{eq:betahat}) oscillates around the minimum of a quadratic potential, $V(\chi)=m^2\chi^2/2$\cite{fedderke/graham/rajendran:2019,BICEPKeck:2021}.

The measured value of $\beta$ suggests that $\chi$ has moved by\cite{minami/komatsu:2020b}
\begin{equation}
\label{eq:interpretation}
\Delta\chi \simeq \frac{10^{-2}}{\alpha}f\,,
\end{equation}
where $\Delta\chi\equiv \chi(\eta_0)-\chi(\eta_{\mathrm{LS}})$. This is sensible: for pion and axion, $\alpha$ is of order the fine-structure constant of EM. If this applies also to $\chi$, $\Delta\chi\simeq f$, which is expected for a cosine potential typical of axionlike fields, $V(\chi)=\mu^4[1+\cos(\varphi/f)]$\cite{freese/frieman/olinto:1990}. See Refs.\cite{fujita/etal:2021a,fujita/etal:2021b,takahashi/yin:2021,mehta/etal:2021,nakagawa/takahashi/yamada:2021,alvey/escudero:2021,choi/etal:2021,obata:prep} for more on interpretation.

\subsection*{Impact of polarised Galactic foreground emission}
We now include the effect of possible foreground $EB$ correlations.
When $C_\ell^{EB,\mathrm{FG}}$ exists, we have the relation $C_\ell^{EB,\mathrm{FG,o}}=\frac12\sin(4\alpha)(C_\ell^{EE,\mathrm{FG}}-C_\ell^{BB,\mathrm{FG}})+C_\ell^{EB,\mathrm{FG}}\cos(4\alpha)$\cite{abitbol/hill/johnson:2016}. We can formally rewrite this as\cite{NPIPE:2022}
\begin{equation}
C_\ell^{EB,\mathrm{FG,o}}=\sqrt{J_\ell^2+\left(C_\ell^{EB,\mathrm{FG}}\right)^2}\sin(4\alpha+4\gamma_\ell)\,,
\end{equation}
where
$J_{\ell}\equiv (C_\ell^{EE,\mathrm{FG}}-C_\ell^{BB,\mathrm{FG}})/2$ and $\tan(4\gamma_\ell)\equiv C_\ell^{EB,\mathrm{FG}}/J_\ell$ is an effective angle for the foreground $EB$. For $C_\ell^{EB,\mathrm{FG}}\propto J_\ell$, $\gamma_\ell$ becomes independent of $\ell$, $\gamma_\ell=\gamma$, and is degenerate with $\alpha$. In this limit and $|\gamma|\ll 1$, the foreground does not yield $\alpha$ but $\alpha+\gamma$, and we measure
$\beta-\gamma$\cite{minami/etal:2019}.

Polarised thermal emission of dust grains is the dominant foreground at the \textit{Planck} HFI frequencies. The \textit{Planck} collaboration also detected intriguing correlations between the dust intensity and $E$- and $B$-mode polarisation, $C_\ell^{TE,\mathrm{dust}}$ and $C_\ell^{TB,\mathrm{dust}}$, respectively\cite{Planck2018XI,weiland/etal:2020}. This is also confirmed by an independent analysis using the distribution of filaments of neutral hydrogen atoms~\cite{clark/etal:2021}.

Detection of $C_\ell^{TB,\mathrm{dust}}$ was not expected. One plausible explanation is based on filaments of hydrogen clouds producing the thermal dust emission and polarisation\cite{huffenberger/rotti/collins:2020}. When the filaments and magnetic field lines are perfectly aligned, a positive $TE$ but no $TB$ or $EB$ correlations are produced. When they misalign by a small angle, $TB$ and $EB$ correlations emerge with the same sign. Such a physical insight results in a model for $\gamma_\ell$ given by\cite{clark/etal:2021,NPIPE:2022}
\begin{equation}
\label{eq:filament}
  \gamma_\ell\simeq
\frac{A_\ell C_\ell^{EE,\mathrm{dust}}}{C_\ell^{EE,\mathrm{dust}}-C_\ell^{BB,\mathrm{dust}}}\frac{C_\ell^{TB,\mathrm{dust}}}{C_\ell^{TE,\mathrm{dust}}}\,,
\end{equation}
where $A_\ell$ is a free amplitude parameter which varies slowly with $\ell$. Using this model to determine $\alpha$, $\beta$ and $\gamma_\ell$ simultaneously, Diego-Palazuelos et al.\cite{NPIPE:2022} find $\beta=0.36^\circ\pm 0.11^\circ$ for nearly full-sky data at the HFI frequencies. The statistical significance exceeds $3\sigma$.

It is reassuring that a frequency-independent $\beta$ is found when adding the LFI data\cite{eskilt:prep}: the Galactic foreground at such low frequencies is no longer dominated by dust emission but is dominated by synchrotron emission. No frequency dependence is therefore consistent with a cosmological signal.

We need to improve our understanding of the foreground polarisation such that we can assign properly the \textit{systematic} uncertainty of the model for $C_\ell^{EB,\mathrm{dust}}$ to the measured value of $\beta$. The hint for cosmic birefringence motivates further work not only on cosmology but also on Galactic science.

We can avoid the issue of the Galactic foreground altogether if we do not rely on it. To this end, we must improve upon the accuracy of calibrating instruments. We comment on this in the `Outlook' section.

\section*{New Physics II: Primordial gravitational waves from the early Universe}
How is it possible that the origin of all structures in the Universe was the quantum-mechanical vacuum fluctuation generated in the early Universe?
Quantum mechanics operates in microscopic atomic worlds, whereas cosmic structures are vastly macroscopic objects. What linked micro- and macroscopic worlds? The leading idea is that the link was provided by a period of accelerated, exponential expansion of space in the early Universe called `cosmic inflation'\cite{guth:1981,sato:1981,linde:1982,albrecht/steinhardt:1982}. See Refs.\cite{steinhardt/turok:2002,nayeri/brandenberger/vafa:2006,brandenberger:2017} for other ideas.
The origin of cosmic structures is therefore explained by a combination of quantum mechanics and general theory of relativity for gravitation -- the two pillars of modern physics\cite{mukhanov:2016}.

According to the idea of inflation, a microscopic wavelength of quantum fluctuations was stretched by enormous expansion of space to become a macroscopic one. Not only was the fluctuation in density (called a scalar mode) generated quantum mechanically\cite{mukhanov/chibisov:1981,hawking:1982,starobinsky:1982,guth/pi:1982,bardeen/turner/steinhardt:1983}, but also the \textit{primordial gravitational wave} (GW; tensor mode) is expected to be produced and stretched to macroscopic wavelengths\cite{grishchuk:1975,starobinsky:1979}. The scalar mode explains the observed properties of cosmic structures\cite{weinberg:2008,peebles:2020}.
The tensor mode is yet to be discovered\cite{tristram/etal:2021,BICEP:2021,SPIDER:2022,tristram/etal:prep}.

\subsection*{Energy density spectrum of GW}
The simplest model of inflation based on a single energy component driving accelerated expansion predicts a stochastic background of primordial GW at all wavelengths, from tens of billions of light-years (comparable to the size of the visible Universe today) to human sizes. More common units used for describing GW are the frequency $f$. A wavelength of tens of billions of light-years corresponds to a frequency of atto Hz ($f=10^{-18}$~Hz). No astrophysical process (e.g., collision of compact objects such as black holes and neutron stars) can generate such a low-frequency GW\cite{caprini/figueroa:2018}; thus, its discovery provides strong evidence for inflation which, in turn, calls for new physics beyond SM\cite{lyth:1997,lyth/riotto:1999}.

\begin{figure*}[ht]
\centering
\includegraphics[width=\linewidth]{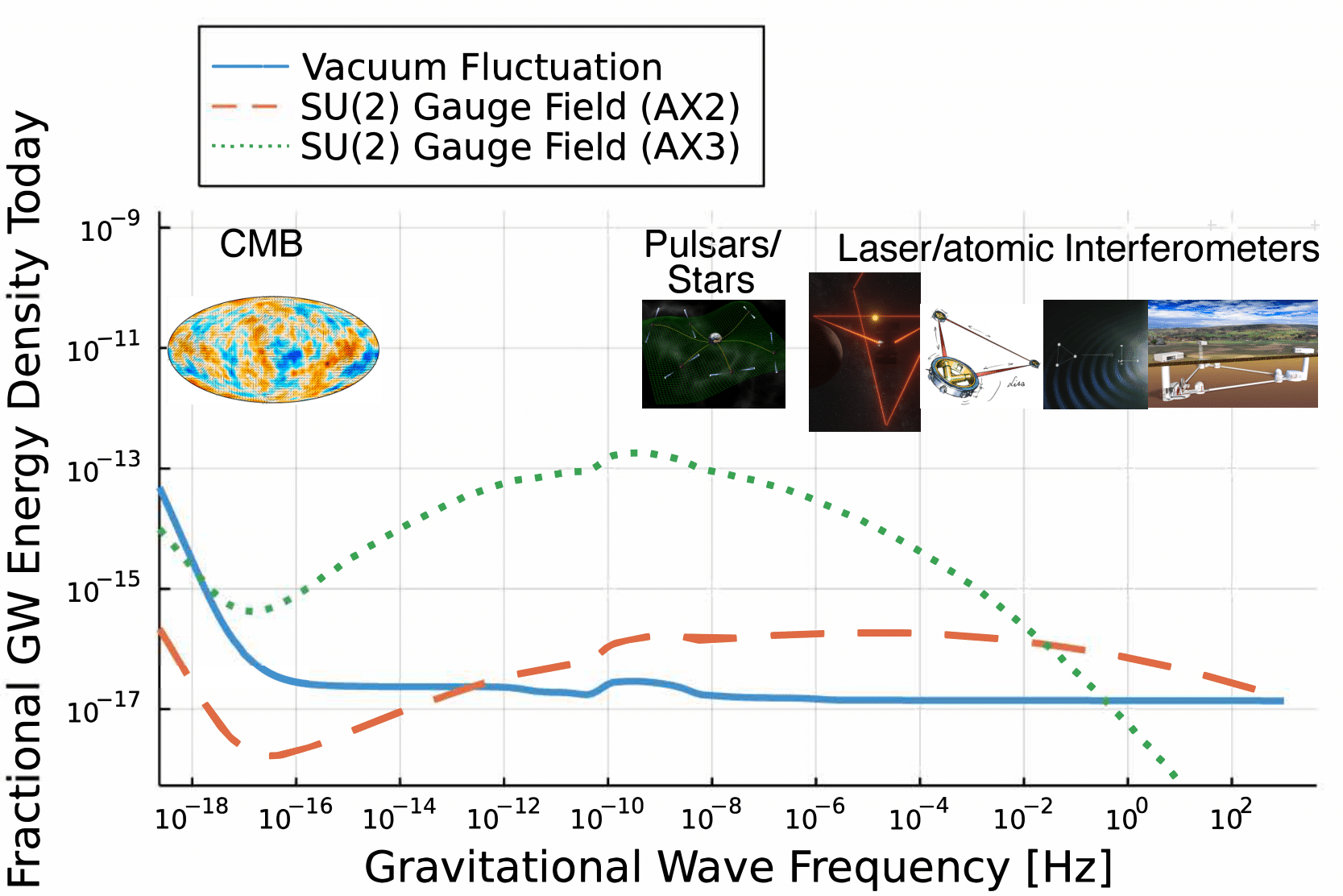}
\caption{{\bf Present-day energy density of primordial GW across 21 decades in frequency.} The vertical axis shows the fractional energy density parameter of GW today, $\Omega_\mathrm{GW}$. The horizontal axis shows the GW frequency in units of Hz. Illustrations of various experimental techniques\cite{bailes/etal:2021} are shown in the corresponding frequency bands. An apparent gap between `CMB' anad `Pulsars/Stars' could be filled by measuring a small departure of the CMB energy spectrum from a Planck spectrum\cite{kite/etal:2021,chluba/etal:2021}.
The solid line shows the vacuum energy contribution of Starobinsky's $R^2$ inflation\cite{starobinsky:1979,starobinsky:1980}.
See Refs.\cite{watanabe/komatsu:2006,saikawa/shirai:2018,kite/etal:2021b} for explanations of the spectral shape. The dashed and dotted lines show $\Omega_\mathrm{GW}$ sourced by SU(2) gauge fields of AX2 and AX3 models given in Ref.\cite{campeti/etal:2021}.}
\label{fig:omega_gw}
\end{figure*}

In Fig.~\ref{fig:omega_gw}, we show typical spectra of the present-day energy density of stochastic GW from inflation discussed in this article. It is common to show the fractional energy density parameter of GW, $\Omega_\mathrm{GW}(f)\equiv \rho_\mathrm{crit}^{-1}d\rho_\mathrm{GW}/d\ln f$, which is the ratio of the GW energy density per logarithmic frequency, $d\rho_\mathrm{GW}/d\ln f$, to the critical density of the Universe, $\rho_\mathrm{crit}$. The most important message of this figure is that the GW spectrum spans a huge range of $f$, and various experiments can measure them across at least 21 decades in $f$\cite{lasky/etal:2016,adshead/etal:2020,campeti/etal:2021,kite/etal:2021,bailes/etal:2021}.

How do we measure GW across 21 decades in $f$? A laser interferometer technique\cite{gertsenshtein/pustovoit:1962,moss/miller/forward:1971} was employed successfully to detect GW from binary black holes\cite{LIGOScientific:2016}, whose wavelength was $\lambda\simeq$~a few thousand km ($f\simeq 10^2$~Hz), comparable to the size of Earth. The upcoming ESA \textit{Laser Interferometer Space Antenna} (\textit{LISA}\cite{LISA:2017}) mission is sensitive to $\lambda\simeq$~a few hundred million km ($10^{-3}$~Hz), comparable to astronomical units. New mission concepts were submitted for
ESA's \textit{Voyage 2050} planning of Large Class science missions in the timeframe 2035-2050, which illustrated various laser and atomic interferometer designs covering $f\simeq 10^{-6}$ to $10$~Hz\cite{sesana/etal:2021,baibhav/etal:2021,sedda/etal:2021,bertoldi/etal:2021}.
We can also measure GW-induced motion of astrophysical bodies such as pulsars\cite{NANOGrav:2020,IPTA:2022} (rotating neutron stars emitting pulses of radio emission) and stars\cite{moore/etal:2017} within our Galaxy to detect GW of $\lambda\simeq$~tens of light-years ($10^{-9}$~Hz). However, none of these techniques can be used to detect GW of $\lambda\simeq$ tens of billions of light-years. We are therefore led to using the whole Universe as a detector: CMB polarisation.

The scalar mode does not produce $B$ modes at linear order in cosmological perturbation\cite{seljak:1997}. This makes $B$ modes a clean probe of the primordial GW\cite{seljak/zaldarriaga:1997,kamionkowski/etal:1997b}. See Supplementary Information for details on how GW produces $B$ modes.

\subsection*{GW from the vacuum fluctuation in spacetime}
Statistical properties of the observed CMB temperature fluctuations and polarisation excited by the scalar mode agree well with the basic prediction of inflation driven by a single energy component\cite{komatsu/etal:2014,Planck2018X}. Observation of the primordial tensor mode would give even stronger evidence for inflation.

What generated tensor modes during inflation? The leading idea is that they were generated by quantum-mechanical vacuum fluctuations in spacetime\cite{grishchuk:1975,starobinsky:1979}.

To set the notation, we first review the power spectrum of tensor modes generated by the vacuum fluctuation. A distance between two events in inhomogeneous spacetime in the presence of tensor modes is given by
\begin{equation}
    ds^2=a^2(\eta)\left[-d\eta^2+\sum_{ij}\exp(h)_{ij}dx^idx^j\right]\,,
\end{equation}
 where $\exp(h)_{ij}=\delta_{ij}+h_{ij}+(1/2)\sum_kh_{ik}h_{kj}+\dots$ is defined in Taylor series. Here, $h_{ij}$ is a $3\times 3$ symmetric real matrix with conditions: (1) $h_{ij}$ does not change a volume (otherwise it induces a change in density like the scalar mode), and (2) $h_{ij}$ is transverse, i.e., perpendicular to the propagation direction, just like an EM wave. The first condition demands the determinant of $\exp(h)_{ij}$ be unity: $\det\left[\exp(h)_{ij}\right]=\exp\left(\sum_{i}h_{ii}\right)=1$; thus, $h_{ij}$ is traceless (1 condition).
The second condition demands $\sum_{i}k^ih_{ij}=0$ where $k^i$ is a wave vector (3 conditions). We are left with $6-1-3=2$ physical degrees of freedom. If we take $k^i$ to be in the $z$ axis, we can write
\begin{equation}
\label{eq:pluscross}
    h_{ij}=\left(\begin{array}{ccc}h_+&h_\times&0\\h_\times&-h_+&0\\0&0&0\end{array}\right)\,.
\end{equation}

Einstein's field equation for $h_{ij}$ at linear order is given by\cite{weinberg:1972}:
\begin{equation}
\label{eq:einstein}
    \Box h_{ij}(t,\bm{x}) = -16\pi G a^{-2}(t) T^\mathrm{t}_{ij}(t,\bm{x})\,,
\end{equation}
where $G$ is the gravitational constant, $\Box\equiv (-g)^{-1/2}\sum_{\mu\nu}\partial/\partial x^\mu\left(\sqrt{-g}g^{\mu\nu}\partial/\partial x^\nu\right)$ the d'Alembert operator for a scalar field in curved spacetime with the Friedmann-Robertson-Walker metric tensor $g_{\mu\nu}$, and $T^\mathrm{t}_{ij}$ the stress-energy source (the superscript `t' stands for tensor modes).

 We define the expansion rate, $H(t)\equiv\dot a/a$, where the dot denotes a time derivative. Integrating this, we find $a(t)\propto \exp[\int H(t')dt']$ and $a(t)$ grows exponentially when $H$ varies slowly with $t$. Accelerated expansion demands $\ddot{a}/a=\dot{H}+H^2>0$, hence $-\dot{H}/H^2<1$. Defining a parameter, $\epsilon_H\equiv -\dot{H}/H^2$, which characterises how slowly $H$ varies, inflation demands $\epsilon_H\ll 1$\cite{lyth/riotto:1999}.

The vacuum fluctuation is a homogeneous solution to equation~(\ref{eq:einstein}) with $T_{ij}^\mathrm{t}=0$.
We calculate variance, $\sum_{ij}\langle h_{ij}^2(t,\bm{x})\rangle$, to quantify the amplitude of tensor modes at a time $t$. The result is divergent because of the short-wavelength contribution, but we consider only the super-Hubble contribution, i.e., variance of a field smoothed over $H^{-1}$. The calculation can be done by computing either a quantum-mechanical vacuum expectation value or an expectation value of a classical random field. They give the same answer:
\begin{align}
    \nonumber
    \sum_{ij}\langle h_{ij}^2(t,\bm{x})\rangle
    &= 2\sum_{p=+,\times}\int\frac{d^3k}{(2\pi)^3}|h_{p,k}(\eta)|^2\\
    &\simeq \int\frac{dk}{k}\frac8{M_{\mathrm{pl}}^2}\left(\frac{H}{2\pi}\right)^2\,,
\end{align}
where\cite{starobinsky:1979}
\begin{equation}
\label{eq:hp-solution}
    h_{p,k}(\eta) = i\frac{H}{M_{\mathrm{pl}}}\frac{e^{-ik\eta}}{k^{3/2}}\left(1+ik\eta\right)\,.
\end{equation}
We used the reduced Planck mass in units of $c=1=\hslash$, $M_{\mathrm{pl}}=1/\sqrt{8\pi G}$. See Supplementary Information for derivation of
equation~(\ref{eq:hp-solution}).

Variance per logarithmic $k$ is independent of $k$, which is known as `scale invariance'. This is the consequence of $H$ being (nearly) constant during inflation, $\epsilon_H\ll 1$. In detail, weak time dependence of $H(t)$ introduces weak $k$ dependence. Variance per logarithmic $k$ is often expressed in terms of the power spectrum ${\cal P}_\mathrm{t}(k)$. It is common to write\cite{weinberg:2008}
\begin{equation}
\label{eq:vacuum-predictions}
    {\cal P}_\mathrm{t}(k)=A_\mathrm{t}\left(\frac{k}{k_*}\right)^{n_\mathrm{t}}\,,\quad
    A_\mathrm{t}=\frac8{M_{\mathrm{pl}}^2}\left(\frac{H(t_*)}{2\pi}\right)^2\,,\quad n_\mathrm{t}=-2\epsilon_H\,,
\end{equation}
where the subscripts `t' stand for tensor modes, $k_*=0.05~\mathrm{Mpc}^{-1}$ is some reference wavenumber (1~Mpc is 3.26 million light-years), and $t_*$ is the time when $k_*=a(t_*)H(t_*)$.

The amplitude of tensor modes is often parametrised by the so-called `tensor-to-scalar ratio' parameter, $r$, defined by $r\equiv A_\mathrm{t}/A_\mathrm{s}$, where $A_\mathrm{s}$ is the corresponding amplitude of scalar modes. The vacuum fluctuation yields $r=16\epsilon_H$\cite{weinberg:2008}. The scalar mode amplitude has been constrained precisely\cite{Planck2018VI}, $A_\mathrm{s}\simeq 2.1\times 10^{-9}$. The upper bound on $r$ from the current $B$-mode data is $r<0.036$~(95\%~C.L.)\cite{BICEP:2021}. This implies an upper bound on the tensor mode amplitude, $A_\mathrm{t}<7.6\times 10^{-11}$, or $H(t_*)<1.9\times 10^{-5}M_{\mathrm{pl}}\simeq 4.7\times 10^{13}$~GeV. Future detection of $r$ at this level implies new physics at an energy scale billion times as large as achieved at CERN's Large Hadron Collider.

The current bound on $r$ has ruled out many models of inflation\cite{komatsu/etal:2014,Planck2018X}. A number of compelling models still remain, including one of the earliest models by Starobinsky\cite{starobinsky:1980} (also see Ref.\cite{nariai/tomita:1971}). His model is based on a quantum correction to the Einstein-Hilbert action, $I=I_\mathrm{GR}+(16\pi G)^{-1}\int d^4x\sqrt{-g}R^2/6M^2$, where $R$ is the Ricci scalar, $M$ is some characteristic energy scale, and $I_\mathrm{GR}=(16\pi G)^{-1}\int d^4x\sqrt{-g}R$ is the the Einstein-Hilbert action. This model is known as `$R^2$ inflation' and predicts $r=12/N^2\simeq 0.0046$ (for $N\simeq 51$), which is one order of magnitude smaller than the current bound.  Here, $N$ is the number of $e$-folds of inflation for which $k_*=a(t_*)H(t_*)$, counted from the end of inflation, i.e., $N=\ln[a(t_{\textrm{end}})/a(t_*)]\simeq H(t_\textrm{end}-t_*)$. This value of $r$ is considered as the next target for CMB experiments.

The same value of $r=12/N^2$ is also predicted\cite{komatsu/futamase:1999} by a model in which a real scalar field driving inflation $\varphi$ is coupled to the Ricci scalar non-minimally, $I=I_\mathrm{GR}+\int d^4x\sqrt{-g}\left[\xi \varphi^2R/2-(\partial\varphi)^2/2-\lambda\varphi^4/4\right]$\cite{futamase/maeda:1987} with $\xi\gg 1$\cite{fakir/unruh:1990} and $(\partial\varphi)^2\equiv\sum_{\mu\nu}g^{\mu\nu}(\partial\varphi/\partial x^\mu)(\partial\varphi/\partial x^\nu)$. This model can be realised when the SM Higgs field is coupled to $R$ non-minimally\cite{bezrukov/shaposhnikov:2008}.
See Refs.\cite{spokoiny:1984,accetta/zoller/turner:1985,fakir/unruh:1990b} for a similar construction in which $I_\mathrm{GR}$ is absent.

After inflation, the primordial GW evolved according to equation~(\ref{eq:einstein}) with $T^\mathrm{t}_{ij}$ provided mostly by free-streaming neutrinos\cite{weinberg:2004}. The resulting energy density spectrum of GW in today's Universe is shown in Fig.~\ref{fig:omega_gw} for $R^2$ inflation. The structures seen in the spectrum reflect various events in the early Universe\cite{watanabe/komatsu:2006,saikawa/shirai:2018,kite/etal:2021b}.

The tensor mode from quantum-mechanical vacuum fluctuations has three important, testable properties.
\begin{itemize}
    \item The power spectrum is nearly scale invariant, with \textit{decreasing} power towards large $k$ with $n_\mathrm{t}=-2\epsilon_H=-r/8<0$.
    \item The probability density function (PDF) is nearly Gaussian\cite{maldacena:2003}.
    \item As the right hand side of equation~(\ref{eq:hp-solution}) does not depend on $p$, the amplitudes of $+$ and $\times$ modes are equal; thus, there is no linear or circular polarisation of the primordial GW. No $C_\ell^{EB}$ is generated.
\end{itemize}
Discovery of violation of any of these properties would lead to a new insight of the fundamental physics behind inflation.

Such observational tests have been performed extensively for scalar modes. We have confirmed that the statistical properties of CMB temperature fluctuations and polarisation from scalar modes are described by Gaussian statistics with high precision\cite{WMAP-NG:2003,komatsu:2010,Planck2018IX}. When writing the scalar-mode power spectrum as ${\cal P}_\mathrm{s}(k)=A_\mathrm{s}(k/k_*)^{n_\mathrm{s}-1}$, we have found that the scalar mode is nearly, but not exactly, scale invariant with $n_\mathrm{s}-1\simeq -0.035$\cite{WMAP:2013b,Planck2018VI}, in agreement with $R^2$ inflation\cite{mukhanov/chibisov:1981}, Higgs inflation\cite{komatsu/futamase:1999,bezrukov/shaposhnikov:2008}, and many others. It is also consistent with $\epsilon_H\ll 1$, as single-field inflation predicts $n_\mathrm{s}-1=-4\epsilon_H-2\delta_H$ with $\delta_H\equiv \ddot{H}/(2H\dot{H})$\cite{weinberg:2008}.

These results strongly support the idea that the observed cosmic structure grew out of quantum vacuum fluctuations generated during inflation. We must now perform the same tests for tensor modes to probe their origin.

\subsection*{Vector field}
In the simplest model of inflation, accelerated expansion is driven by a single energy component such as a scalar field $\varphi$\cite{lyth/riotto:1999}. This does not mean that there were no other matter fields during inflation; on the contrary, matter fields must have existed to allow the energy density of $\varphi$ to be converted eventually to a plasma of SM particles after inflation. This process is called `reheating'~\cite{lozanov:2019}. In cosmology, we often call massive and massless particles `matter' and `radiation', respectively. We do not make such a distinction here and call all fields other than $\varphi$ `matter fields' regardless of the mass of particles.

Matter fields can source tensor modes via the stress-energy tensor, $T^\mathrm{t}_{ij}$, on the right hand side of Einstein's field equation~(\ref{eq:einstein}). The vacuum fluctuation is the homogeneous solution of this equation. We now discuss inhomogeneous solutions sourced by matter fields.

To calculate $T_{ij}$, we need a Lagrangian density of matter fields, ${\cal L}_\mathrm{m}$, in the action
\begin{equation}
    I=I_\mathrm{GR}+I_{\mathrm{inf},\varphi}+\int d^4x\sqrt{-g}{\cal L}_\mathrm{m}\,,
\end{equation}
where $I_{\mathrm{inf},\varphi}$ is the action for $\varphi$ driving inflation, e.g., $I_{\mathrm{inf},\varphi}=\int d^4x\sqrt{-g}\left[-(\partial\varphi)^2/2-V(\varphi)\right]$ where $V(\varphi)$ is $\varphi$'s potential. The stress-energy tensor is given by
\begin{equation}
    T_{ij}^\mathrm{m} = \frac{-2}{\sqrt{-g}}\frac{\delta(\sqrt{-g}{\cal L}_\mathrm{m})}{\delta g^{ij}}\,.
\end{equation}
To extract $T_{ij}^\mathrm{t}$, which is transverse and traceless, we remove $g_{ij}T^\mathrm{m}/3$ from $T_{ij}^\mathrm{m}$ where $T^\mathrm{m} = \sum_{ij}g^{ij}T_{ij}^\mathrm{m}$ is trace, and extract a component transverse to the wave vector in Fourier space.

For example, consider a massless free vector field, $A_\mu$, like an EM field we discussed in the earlier sections on cosmic birefringence. The traceless component relevant for tensor modes is $T_{ij}^A-g_{ij}T^A/3=-a^2(E_iE_j+B_iB_j)+g_{ij}(\bm{E}\cdot\bm{E}+\bm{B}\cdot\bm{B})/3$.
It is important to realise that this is second order in perturbation when geometry of the background spacetime is isotropic. If a homogeneous vector field with a preferred direction exists, it breaks isotropy globally at the background level; thus, $\bm{E}$ and $\bm{B}$ must be a perturbation.

This result is generic. The so-called scalar-vector-tensor decomposition theorem states that scalar, vector, and tensor modes are decoupled and evolve independently of each other at first order in perturbation, when the background spacetime is homogeneous and isotropic\cite{kodama/saaski:1984}. Therefore, not only vectors but also scalar matter fields can source GW only at second order\cite{tomita:1967,matarrese/mollerach/bruni:1998}. There is a guaranteed lower bound for the amplitude of a stochastic background of GW sourced by the second-order scalar mode in the early Universe\cite{ananda/clarkson/wands:2007,baumann/etal:2007}. Additional scalar degrees of freedom during inflation\cite{carney/etal:2012,cook/sorbo:2012,biagetti/fasiello/riotto:2013,senatore/silverstein/zaldarriaga:2014,cai/etal:2021} and other scenarios\cite{martineau/brandenberger:2008,brandenberger/etal:2007a,brandenberger/etal:2007b} can enhance the amplitude of GW. We refer the readers to a recent review article by Dom\`enech\cite{domenech:2021} for details on this rich subject.

Can we still find an interesting level of tensor modes from a massless free vector field? The answer is no, if all we have is a massless free field. As we saw already, a massless free vector field is conformally coupled to gravitation, with the equation of motion $A_i''+k^2A_i=0$; thus, $A_i$ is oscillatory, and the EM fields decay as $\bm{E}\propto a^{-2}$ and $\bm{B}\propto a^{-2}$.

The situation changes dramatically when we include the
CS term given in equation~(\ref{eq:CS}). What is $\chi$ in this context? It can be the field driving inflation with an action
$I=I_\mathrm{GR}+I_{\mathrm{inf},\chi}+I_A+I_\mathrm{CS}$\cite{anber/sorbo:2010} or an additional `spectator' field whose energy density is negligible compared to $\varphi$ with an action $I=I_\mathrm{GR}+I_{\mathrm{inf},\varphi}+I_\chi+I_A+I_\mathrm{CS}$\cite{barnaby/etal:2012}. Here, $\bm{E}$ and $\bm{B}$ need not be the SM EM fields; most likely they are not because of many theoretical challenges and constraints on generation of astrophysically interesting EM fields during inflation\cite{ratra:1992,demozzi/mukhanov/rubinstein:2009,barnaby/namba/peloso:2012,maleknejad/sheikh-jabbari/soda:2013,kobayashi/afshordi:2014}. We thus assume that $\bm{E}$ and $\bm{B}$ are some new vector fields present during inflation.

The equation of motion for $A_i$ is given by equation~(\ref{eq:vectorequation}).
As the scale factor is given by $a(\eta)\simeq -(H\eta)^{-1}$ for $-\infty<\eta<0$ during inflation, we obtain
\begin{equation}
\label{eq:vectorequation2}
    A''_{\pm}+\left[k^2\mp\frac{2k\xi}{(-\eta)}\right]A_{\pm}=0\,,
\end{equation}
where $\xi\equiv \alpha\dot{\chi}/(2fH)$\cite{anber/sorbo:2010}.
We take $\alpha\dot{\chi}>0$ (hence $\xi>0$) without loss of generality. Then, the effective frequency squared, $\omega_\pm^2=k^2\mp 2k\xi/(-\eta)$, can become negative for the $+$ state, and the mode function is amplified during inflation. The exact solution that matches a positive frequency mode deep in the short-wavelength regime, $k|\eta|\gg 1$, can be obtained
when $\xi$ varies slowly with $t$, and is given by Whittaker's function\cite{barnaby/pajer/peloso:2012}.
The amplified vector field sources tensor modes via $\Box h_{ij}=16\pi G(E_iE_j+B_iB_j)^\mathrm{TT}$, where the subscript `TT' indicates a transverse and traceless component. As the tensor mode is sourced by product of two perturbations, it is highly non-Gaussian\cite{anber/sorbo:2012,namba/etal:2016,shiraishi/etal:2016}.

The resulting power spectrum is \textit{chiral}, i.e., the amplitudes of right- and left-handed polarisation states of GW are different. For GW propagating in the $z$ axis, we define right- and left-handed states (helicity $+2$ and $-2$ respectively) as $h_\mathrm{R}=(h_+-ih_\times)/\sqrt{2}$ and $h_\mathrm{L}=(h_++ih_\times)/\sqrt{2}$. The corresponding power spectra are given by\cite{sorbo:2011,barnaby/namba/peloso:2011}
\begin{align}
\label{eq:PRu1}
{\cal P}_\mathrm{t,R} &\simeq \frac{4}{M_\mathrm{pl}^2}\left(\frac{H}{2\pi}\right)^2\left[1+8.6\times 10^{-7}\frac{H^2}{M_\mathrm{pl}^2}\frac{e^{4\pi\xi}}{\xi^6}\right]\,,\\
\label{eq:PLu1}
{\cal P}_\mathrm{t,L} &\simeq \frac{4}{M_\mathrm{pl}^2}\left(\frac{H}{2\pi}\right)^2\left[1+1.8\times 10^{-9}\frac{H^2}{M_\mathrm{pl}^2}\frac{e^{4\pi\xi}}{\xi^6}\right]\,.
\end{align}
The first term in the square bracket is the vacuum contribution. Adding them reproduces the vacuum result given in equation~(\ref{eq:vacuum-predictions}): ${\cal P}^{\rm vac}_\mathrm{t,R}+{\cal P}^{\rm vac}_\mathrm{t,L}= (8/M_\mathrm{pl}^2)(H/2\pi)^2$. The second terms differ by more than two orders of magnitude, i.e., chiral GW. We can probe this by measuring the $EB$ power spectrum of CMB polarisation data\cite{lue/wang/kamionkowski:1999,sorbo:2011} as well as circular polarisation of GW using laser interferometers\cite{seto:2006,seto/taruya:2007,crowder/etal:2012}.

As $B$ modes from primordial GW are yet to be found, the current $EB$ data do not constrain the parameter space of this model. If GW were found, the lack of detection of  $EB$ would constrain the parameter space; however, if GW were not found, the lack of $EB$ would not mean much. Conversely, if GW were not found but $EB$ were found, it would point towards a different mechanism for $EB$, such as cosmic birefringence.

For the vacuum contribution, the scale-dependence of ${\cal P}_\mathrm{t}$ was given by weak time dependence of $H(t)$, resulting in the decreasing power at large $k$ with $n_\mathrm{t}=-2\epsilon_H<0$. For the sourced contribution, additional scale-dependence is given by weak time dependence of $\xi(t)\propto \dot{\chi}/H$. Depending on the potential for $\chi$, $\dot{\chi}/H$ can increase during inflation ($\chi$ speeds up as it rolls down on its potential $V(\chi)$), which results in \textit{increasing} power at large $k$\cite{cook/sorbo:2012,barnaby/pajer/peloso:2012}.

The amplified vector field sources not only tensor modes, but also scalar modes via $\Box\chi-\partial V/\partial\chi=-(\alpha/f)\bm{E}\cdot\bm{B}$\cite{anber/sorbo:2010}. This scalar mode is highly non-Gaussian\cite{barnaby/peloso:2011}. As there is no evidence for scalar-mode non-Gaussianity in the CMB data\cite{WMAP-NG:2003,komatsu:2010,Planck2018IX}, the model producing a sizable primordial GW in the scales relevant to CMB is highly constrained if $\chi$ is the field driving inflation\cite{barnaby/peloso:2011,barnaby/namba/peloso:2011}. The constraint is lessened when $\chi$ is an additional spectator field coupled to $\varphi$ only gravitationally\cite{barnaby/etal:2012,cook/sorbo:2013,namba/etal:2016}. A significant amount of GW can still be produced at interferometer scales (Fig.~\ref{fig:omega_gw}) without violating the CMB constraint on scalar non-Gaussianity\cite{barnaby/pajer/peloso:2012,bartolo/etal:2016}.

To summarise, the tensor mode sourced by the vector field, which is amplified by the CS term, violates \textit{all} of the three properties of the vacuum contribution: it produces non-scale-invariant, non-Gaussian, and chiral primordial GW. Such observational tests open a completely new window into the particle physics of inflation.

\subsection*{Non-Abelian gauge field}
Scalar and vector matter fields can source tensor modes only at second order, when the background spacetime is homogeneous and isotropic. As there is no evidence for statistical anisotropy in the \textit{Planck} data\cite{kim/komatsu:2013}, we assume isotropic geometry. Is there a matter field which can source tensor modes at linear order without breaking isotropy?

The answer is yes: it is a non-Abelian gauge field with SU(2) symmetry, $\bm{A}_\mu$. The vector field example in the previous section can be regarded as an Abelian gauge field with U(1) symmetry.

In 2011, Maleknejad and Sheikh-Jabbari discovered a new class of homogeneous and isotropic inflationary solutions with $\bm{A}_\mu$, which they called `Gauge-flation'\cite{maleknejad/sheikh-jabbari:2011,maleknejad/sheikh-jabbari:2013}. As a massless free gauge field is conformally coupled, they broke conformal invariance by adding
a term ${\cal L}_\mathrm{Gf}=\kappa(F\tilde F)^2/96$ to the Lagrangian density ($\kappa$ is a constant). Here, the CS term is
\begin{equation}
   F\tilde F= \sum_{a=1}^3\sum_{\mu\nu}F^a_{\mu\nu}\sum_{\mu'\nu'}\frac{\epsilon^{\mu\nu\mu'\nu'}}{2\sqrt{-g}}F^a_{\mu'\nu'}\,,
\end{equation}
where
\begin{equation}
F^a_{\mu\nu}= \frac{\partial A^a_\nu}{\partial x^\mu}-\frac{\partial A^a_\mu}{\partial x^\nu}+g_A\sum_{b=1}^3\sum_{c=1}^3\epsilon^{abc}A_\mu^bA_\nu^c\,,
\end{equation}
with $\bm{A}_\mu= \sum_a A^a_\mu \bm{\sigma}_a/2$ and $\bm{\sigma}_a$ ($a=1,2,3$) being the Pauli matrices. The self-coupling constant is $g_A$.

The Maleknejad-Sheikh-Jabbari solution is characterised by a homogeneous and isotropic vacuum expectation value for $A_i^a=a(t)Q(t)\delta_i^a$ and $A_0^a\equiv 0$. Imagine three vectors: $A_i^1$, $A_i^2$ and $A_i^3$, which align with the three axes of Cartesian coordinates such that there is no preferred direction. This configuration can be rotated by an arbitrary 3-rotation, breaking isotropy. However, we can always find a global SU(2) transformation which cancels the 3-rotation, restoring isotropy. The fundamental reason for this behaviour is that the adjoint representation of SU(2) is isomorphic to SO(3), i.e., symmetry group of 3-rotation.

In 2012, Adshead and Wyman found the same result using another action, which they called `Chromo-natural inflation'\cite{adshead/wyman:2012},
$I=I_\mathrm{GR}+I_{\mathrm{inf},\chi}+I_A+I_\mathrm{CS}$, with a cosine potential typical for axionlike fields, $V(\chi)=\mu^4[1+\cos(\varphi/f)]$\cite{freese/frieman/olinto:1990}. This action is similar to that of the vector field (Abelian gauge field) example discussed in the previous section but with $F^2$ and $F\tilde F$ computed for SU(2). It was then found that the Gauge-flation action with $\kappa=3\alpha^2/\mu^4$
could be obtained from the Chromo-natural one by integrating out $\chi$ near the minimum of its potential\cite{adshead/wyman:2012b,sheikh-jabbari:2012}.
They therefore share similar phenomenology\cite{maleknejad/sheikh-jabbari/soda:2013}. Possible embedding of this model in a more fundamental theory is discussed in Refs.\cite{dallagata:2018,mcdonough/alexander:2018,agrawal/fan/reece:2018,holland/zavala/tasinato:2020}.

The SM weak interaction acts on left-handed fermions. Recently, Maleknejad embedded the above SU(2) field in the right-handed sector of weak interaction in the so-called left-right symmetric extension of SM\cite{maleknejad:2020b,maleknejad:2021}. Her model has a rich phenomenology not only for the primordial GW from inflation but also for dark matter and the origin of matter-anti matter asymmetry.

The isotropic solution of Gauge-flation/Chromo-natural inflation is an attractor: even if spacetime geometry of the Universe was highly anisotropic initially, it approaches the isotropic solution\cite{maleknejad/erfani:2014,wolfson/maleknejad/komatsu:2020,wolfson/etal:2021}. Isotropy is broken when $A_i^a$ is allowed to acquire different mass $m_a$ depending on $a$. When a mass term is added to the Lagrangian density, $m^2_a A_\mu^a A_\nu^a g^{\mu\nu}/2$, anisotropic solutions can exist for $m_1\neq m_2=m_3$\cite{adshead/liu:2018}.

Expanding $A_\mu^a$ around the vacuum expectation value, we write $A_i^a(t,\bm{x})=a(t)Q(t)\delta_i^a + \delta A_i^a(t,\bm{x})$. The perturbation contains 1 tensor, 2 vector, and 3 scalar modes, $\delta A_i^a=t_{ai}+\dots$, where $t_{ai}$ is transverse and traceless (tensor mode) and `$\dots$' contains scalar and vector modes\cite{maleknejad/sheikh-jabbari:2011,maleknejad/sheikh-jabbari:2013}. The stress-energy tensor consists of the background and perturbed parts, $T_{ij}^A=\bar T_{ij}^A+\delta T_{ij}^A$. There was no background part in the Abelian case. The perturbed stress-energy tensor for $t_{ai}$ is
\begin{align}
\nonumber
    \delta T_{ij}^\mathrm{t} = -&\frac2a\frac{d(aQ)}{dt}t_{ij}'
    +2g_AQ^2\Big\{g_A(aQ)t_{ij}\\
    &\left.-\frac12\left[\sum_{ab}\epsilon^{iba}\frac{\partial t_{aj}}{\partial x^b}+(i\leftrightarrow j)\right]\right\}\,.
\end{align}
This is linear in $t_{ij}$, as promised.

When tensor modes propagate in the $z$ axis, we can define $+$ and $\times$ modes in the same way as for GW (equation~(\ref{eq:pluscross})) as well as left- and right-handed states, $t_L=(t_++it_\times)/\sqrt{2}$ and $t_R=(t_+-it_\times)/\sqrt{2}$. Then, we find the stress-energy tensor for $t_{R}$ and $t_L$ in Fourier space as
\begin{equation}
\delta T^\mathrm{t}_{R/L}=-\frac2a\frac{d(aQ)}{dt}t_{R/L}'
    +2g_AQ^2\left(g_AaQt_{R/L}\mp k_3t_{R/L}\right)\,,
\end{equation}
where the minus and plus signs are for the $R$ and $L$ modes, respectively.
Symmetry permits us to generalise this result for arbitrary propagating directions by replacing $k_3$ with $k$.
Right/left-handed GW are sourced linearly by $t_{R/L}$ in $\delta T^\mathrm{t}_{R/L}$, respectively.

The equations of motion for $t_R$ and $t_L$ are\cite{adshead/martinec/wyman:2013,dimastrogiovanni/peloso:2012,maleknejad/sheikh-jabbari/soda:2013}
\begin{equation}
    t_{R/L}''+\left[k^2+\frac2{\eta^2}\left(m_Q\xi\mp (-k\eta)(m_Q+\xi)\right)\right]t_{R/L}={\cal O}(h_{R/L})\,,
\end{equation}
where the minus and plus signs are for the $R$ and $L$ modes, respectively. Here, $\xi=\alpha\dot{\chi}/(2fH)$ is the same as in the Abelian case, while $m_Q\equiv g_AQ/H$ is unique to the SU(2) case because it has the self-coupling constant $g_A$. The right hand side contains $h_{L/R}$, which we can ignore. Notice the minus sign in front of $-k\eta$ for $t_R$; this signals amplification again, just like for the Abelian case.
For $\xi>0$, $t_R$ is amplified when $\sqrt{2}(-1+\sqrt{2})m_Q<-k\eta<\sqrt{2}(1+\sqrt{2})m_Q$, i.e., $0.6m_Q<-k\eta<3.6m_Q$. The solution that matches a positive frequency mode deep in the
short-wavelength regime is given by Whittaker's function when $\xi$ and $m_Q$ vary slowly.

Just like for the Abelian case given in equations~(\ref{eq:PRu1}) and (\ref{eq:PLu1}), the resulting power spectrum is chiral:
\begin{equation}
{\cal P}_\mathrm{t,R/L} \simeq \frac{4}{M_\mathrm{pl}^2}\left(\frac{H}{2\pi}\right)^2\left[1+
\frac{Q^2}{2M_\mathrm{pl}^2}|{\cal G}_{R/L}(m_Q)|^2e^{\pi(m_Q+\xi)}\right]\,,
\end{equation}
where ${\cal G}_{R/L}$ is given in Appendix E of Ref.\cite{maleknejad/komatsu:2019}. It suffices to say $|{\cal G}_{R}|^2\gg |{\cal G}_{L}|^2$ for $\xi>0$.

During inflation $\xi\simeq m_Q+m_Q^{-1}$ and a successful phenomenology requires $\sqrt{2}<m_Q\lesssim\mathrm{a~few}$\cite{dimastrogiovanni/peloso:2012,adshead/martinec/wyman:2013b}. The basic phenomenology is similar to the Abelian case discussed earlier: depending on the potential for $\chi$, weak time dependence of $m_Q$ and $\xi$
can make ${\cal P}_\mathrm{t}$ increase towards large $k$. The sourced GW is chiral, yielding non-vanishing $C_\ell^{EB}$ for CMB experiments and circularly polarised GW for interferometers\cite{thorne/etal:2018}.
For the Abelian case, the sourced GW was highly non-Gaussian because it was sourced at second order in EM fields. For the SU(2) case, it is highly non-Gaussian because of the self-coupling of gauge fields\cite{agrawal/fujita/komatsu:2018,agrawal/fujita/komatsu:2018b,dimastrogiovanni/etal:2018,fujita/namba/obata:2019,fujita/etal:2022}. Thus, GW sourced by the SU(2) gauge field again violates all of the three properties of the vacuum contribution. This makes a strong case for observational tests of these properties, once the primordial GW is discovered in $B$-mode polarisation or other techniques shown in Fig.~\ref{fig:omega_gw}.

The action may contain the so-called `gravitational Chern-Simons (GCS) term', $\chi R\tilde R$\cite{alexander/yunes:2009}, which also induces parity-violating correlations in the CMB data\cite{lue/wang/kamionkowski:1999,saito/ichiki/taruya:2007,contaldi/magueijo/smolin:2008}. When considered simultaneously with $I_{\mathrm{CS}}$, the impact of GCS is minor\cite{mirzagholi/etal:2020}, justifying our ignoring this term in this section.

Production of GW by the SU(2) gauge field is so efficient that the original Gauge-flation and Chromo-natural inflation models have been ruled out by upper bounds on $B$ modes\cite{adshead/martinec/wyman:2013b,namba/dimastrogiovanni/peloso:2013}. The model can be made compatible with observational data by modifying $\chi$'s potential\cite{maleknejad:2016,caldwell/devulder:2018} or kinetic term\cite{watanabe/komatsu:prep}; making $\chi$ a spectator field\cite{dimastrogiovanni/fasiello/fujita:2016,iarygina/sfakianakis:2021}; giving mass to the gauge field via a Higgs-like mechanism\cite{nieto/rodriguez:2016,adshead/etal:2016,adshead/sfakianakis:2017}; delaying the onset of the Chromo-natural inflation phase so that GW is not produced at the CMB scale but at higher frequencies\cite{obata/miura/soda:2015,obata/soda:2016,domcke/etal:2019b}, etc.

Here we show ${\cal P}_\mathrm{t}$ of the spectator model\cite{dimastrogiovanni/fasiello/fujita:2016} with an action $I=I_\mathrm{GR}+I_{\mathrm{inf},\varphi}+I_{\chi}+I_A+I_\mathrm{CS}$. The detailed shape of ${\cal P}_\mathrm{t}$ is determined by time evolution of $m_Q$ and $\xi$. For a cosine potential, $\chi$ speeds up initially, reaches the maximum velocity at the inflection point of the potential $\chi(t_*)=\pi f/2$, and slows down. The sourced tensor-mode power spectrum has a peak at the wavenumber $k_\mathrm{p}$ that corresponds to $\chi(t_*)$\cite{thorne/etal:2018}
\begin{equation}
    {\cal P}^{\mathrm{source}}_{\mathrm{t,R}}(k) =
    r_* {\cal P}_{\mathrm{s}}(k)
    \exp\left[-\frac1{2\Sigma^2}\ln^2\left(\frac{k}{k_{\mathrm{p}}}\right)\right]\,,
\end{equation}
where $r_*$ is the tensor-to-scalar ratio at $k=k_{\mathrm{p}}$, ${\cal P}_{\mathrm{s}}(k)$ is the scalar-mode power spectrum, and $\Sigma$ characterises the width of the peak. Other shapes are possible for other forms of $V(\chi)$\cite{fujita/sfakianakis/shiraishi:2019}.
In terms of the model parameters, we write $m_*\equiv m_Q(t_*)=(g_A^2\mu^4/3\alpha H^4)^{1/3}$, $\xi_*\simeq m_*+m_*^{-1}$, $\Sigma^2=(\alpha/2\xi_*)^2/[2{\cal I}(m_*)]$, and ${\cal I}(m_*)\simeq 0.666+0.81m_*-0.0145m_*^2-0.0064m_*^3$.

In Fig.~\ref{fig:omega_gw}, we show $\Omega_{\mathrm{GW}}$ computed from the `AX2' ($r_*=0.15$, $k_{\mathrm{p}}=10^{11}~\mathrm{Mpc}^{-1}$, $\Sigma=8$) and `AX3' ($r_*=50$, $k_{\mathrm{p}}=10^{6}~\mathrm{Mpc}^{-1}$, $\Sigma=4.8$) models given in Ref.\cite{campeti/etal:2021}. We use ${\cal P}_{\mathrm{s}}=A_\mathrm{s}(k/k_*)^{n_\mathrm{s}-1}$ but other forms are possible at larger $k$ than constrained precisely by CMB observations. The shapes of $\Omega_{\mathrm{GW}}$ are completely different from that of the vacuum contribution and distinguishable
not only in the CMB polarisation data sets\cite{LiteBIRD:2022}, but also in the other data sets\cite{campeti/etal:2021}. Measuring the shape of $\Omega_{\mathrm{GW}}$ across 21 decades in frequency promises a breakthrough in our understanding of the fundamental physics behind inflation.

\begin{figure*}[ht]
\centering
\includegraphics[width=\linewidth]{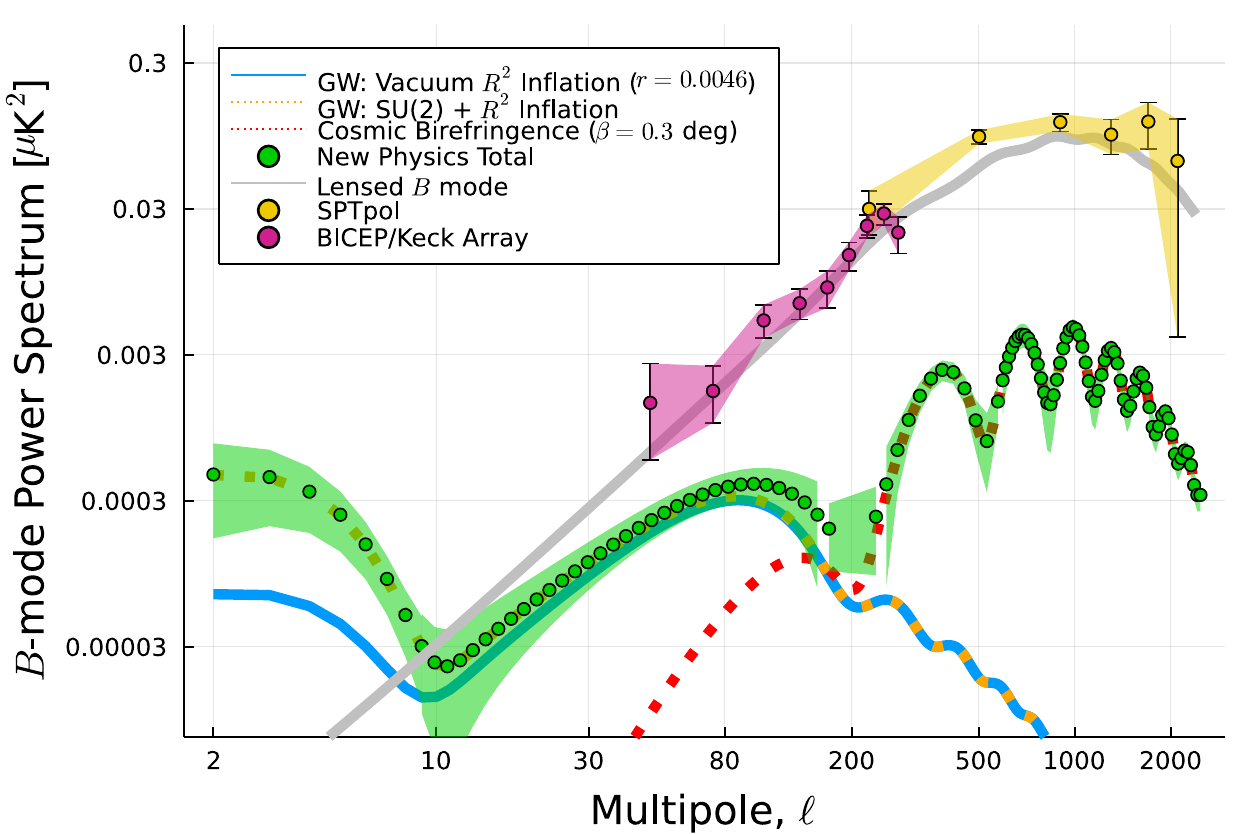}
\caption{{\bf $B$-mode polarisation power spectra.}
We show $\ell(\ell+1)C_\ell^{BB}/2\pi$ in units of $\mu\mathrm{K}^2$ from cosmic birefringence with $\beta=0.3^\circ$ (red dashed) and a sourced GW from SU(2) gauge fields added to the vacuum contribution of $R^2$ inflation (orange dashed). The green filled circles show the sum of these contributions, which are averaged over multipoles in bins centered at $\ell$ as shown. The lensed $B$-mode power spectrum (grey solid) is subtracted. The bands indicate the 68\%~C.L. intervals assuming full-sky and noiseless experiments (the lensed $B$-mode contribution is included in the intervals). The blue solid line shows the vacuum contribution alone. The data points of SPTpol\cite{sayre/etal:2020} and BICEP/Keck Array\cite{BICEP:2021} are the same as in Fig.~\ref{fig:cl}.}
\label{fig:requirement}
\end{figure*}

In Fig.~\ref{fig:requirement}, we show
the $B$-mode power spectrum computed from another choice of axion-SU(2) parameters: $r_*=0.023$, $k_{\mathrm{p}}=3.4\times 10^{-4}~\mathrm{Mpc}^{-1}$, and $\Sigma=1.1$ (P. Campeti, private communication), which is added to the vacuum contribution of $R^2$ inflation. This parameter was chosen such that the sourced GW enhances the reionisation bump at $\ell\lesssim 10$ significantly compared to the vacuum contribution, while satisfying
constraints from backreaction discussed below. We also show the 68\%~C.L. intervals of ideal experiments with full-sky coverage and noise much below the lensed $B$ modes (i.e., $\ll 5~\mu$K~arcmin). In this case, they can be treated effectively as noiseless experiments. We find that the sourced contribution is clearly detectable over the vacuum contribution, if we have access to low multipoles, i.e., full-sky coverage by a satellite mission such as \textit{LiteBIRD}. While no foreground contamination or instrumental systematics was included in the uncertainty here, more detailed analysis including them reached the same conclusion\cite{LiteBIRD:2022}.

Observations of non-Gaussianity of $B$ modes would not only be revolutionary, but also place useful constraints on the parameter space of the sourced GW\cite{agrawal/fujita/komatsu:2018,agrawal/fujita/komatsu:2018b}. The prospects for detecting $EB$ from chiral GW used to be good when a large $r$ was allowed\cite{saito/ichiki/taruya:2007,thorne/etal:2018}. While the current upper bound on $r$ from the recombination bump at $\ell\simeq 80$ (Fig.~\ref{fig:cl})\cite{BICEP:2021} is making this measurement a challenge, there is still a room for a detectable $EB$ signal in the reionisation bump.

In Fig.~\ref{fig:requirement}, we also show the $B$-mode power spectrum from cosmic birefringence, $C_\ell^{BB,\mathrm{o}}=\sin^2(2\beta)C_\ell^{EE}$, with $\beta=0.3^\circ$. This contribution dominates at $\ell\gtrsim 200$, which is accessible to ground-based observatories such as CMB Stage-4. The synergy of space-borne and ground-based experiments in the quest for new physics is obvious.

The presence of a strong EM field creates pairs of electrons and positrons via the so-called Schwinger process\cite{schwinger:1951}. Similar phenomena occur here: the background $\chi$ and gauge fields create a copious amount of particles coupled to either or both of them, such as charged scalar\cite{lozanov/maleknejad/komatsu:2019}, massless\cite{domcke/etal:2019} and massive\cite{mirzagholi/makeknejad/lozanov:2020,maleknejad:2020} fermions, and spin-2 particles\cite{maleknejad/komatsu:2019}. When the energy and momentum densities of the produced particles become large, they give `backreaction' to the background equations of motion for $\chi$ and gauge fields, potentially spoiling phenomenological success of the model\cite{dimastrogiovanni/fasiello/fujita:2016,fujita/namba/tada:2018}.

Ishiwata et al.\cite{ishiwata/komatsu/obata:2022} analysed the backreaction of spin-2 particles in detail, finding that the sourced GW can exceed the vacuum contribution by many orders of magnitude without spoiling dynamics of the background $\chi$ and gauge fields. The enhancement of GW at high frequencies shown in Fig.~\ref{fig:omega_gw} is therefore allowed.

In the GW frequency measured by CMB experiments, the additional constraints on $P_\mathrm{s}\simeq 2.1\times 10^{-9}$ and $r<0.036$ allow the sourced GW to exceed the vacuum contribution by more than an order of magnitude.

While scalar modes in $\delta A_i^a$ are not excited for $m_Q>\sqrt{2}$\cite{dimastrogiovanni/peloso:2012,adshead/martinec/wyman:2013b}, tensor modes can source scalar modes at second order\cite{papageorgiou/peloso/unal:2018,papageorgiou/peloso/unal:2019}. Such non-linearly sourced scalar modes not only modify $P_\mathrm{s}(k)$ but also add non-Gaussian fluctuations which are strongly constrained by the \textit{Planck} temperature anisotropy data\cite{Planck2018IX}.
This effect would constrain viable parameter space of the model in the GW frequency measured by CMB experiments further. The detailed analysis is yet to be done, though the sourced and vacuum contributions can still be comparable. In particular, the former can
exceed the latter by a factor of 5 at low multipoles, $\ell\lesssim 10$ where the constraint on the scalar mode is not strong\cite{ishiwata/komatsu/obata:2022}. The $B$-mode power spectrum at such low multipoles will be measured by the \textit{LiteBIRD} mission\cite{LiteBIRD:2022}.

The above phenomenology is valid when extended to SU($N$), provided that it contains SU(2) as a subgroup\cite{caldwell/devulder:2018,holland/zavala/tasinato:2020,fujita/etal:prep}. Can $\bm{A}_\mu$ be the SM SU(2) gauge field? It is possible, though $\chi$ coupled to the CS term is beyond SM. It is also possible that $\chi$ couples to the SM particles, as in the Peccei-Quinn solution to the strong \textit{CP} problem in QCD\cite{peccei/quinn:1977,weinberg:1978,wilczek:1978}.

\section*{Outlook}
Sensitivity of polarisation to new physics that violates parity symmetry offers exciting opportunities for discovery, which may tell us the fundamental physics behind dark matter, dark energy, and cosmic inflation. The science topics described in this article may shed new light on the way polarisation data from on-going\cite{adachi/etal:2020,adachi/etal:2020b,aiola/etal:2020,sayre/etal:2020,dutcher/:2021,BICEP:2021,SPIDER:2022,dahal/etal:2022} and future CMB experiments\cite{SimonsObservatory:2019,SPO:2020,CMB-S4:2019,LiteBIRD:2022} are obtained, calibrated, and analysed.

Searching for new physics requires new strategy. Efforts to discover the $B$-mode power spectrum from primordial GW are endorsed strongly by American National Academies's \textit{Decadal Survey on Astronomy and Astrophysics 2020} (Astro2020), and currently drive requirements for design of future experiments\cite{kamionkowski/kovetz:2016}. However, do they also satisfy requirements for the $EB$ science and tests of Gaussianity of $B$ modes?

Miscalibration of polarisation-sensitive orientations of detectors on the focal plane of a telescope with respect to the sky as well as cross-polarisation coupling induced by systematics of beam \cite{hu/hedman/zaldarriaga:2003,rosset/etal:2007,odea/etal:2007,shimon/etal:2008} and optical components such as half-wave plates\cite{bao/etal:2012,duivenvoorden/etal:2021,giardiello/etal:2021} yield $\alpha$, hence a spurious $B$-mode power spectrum, $C_\ell^{BB,\mathrm{o}}=\sin^2(2\alpha)C_\ell^{EE}$.
Reducing this to sufficiently small level required for the $B$-mode science sets a requirement for $\alpha$, which is below $0.2^\circ$\cite{abitbol/etal:2021} or even below $0.1^\circ$\cite{vielva/etal:prep} (see Fig.~\ref{fig:requirement} for $B$ modes of $0.3^\circ$). Achieving this in practice is a significant challenge, and the CMB community often calibrates $\alpha$ using $C_\ell^{EB,\mathrm{o}}$, assuming null cosmological signals\cite{keating/shimon/yadav:2012,RAC:2022}. This obviously compromises our ability to search for new parity-violating physics, which has a hint of $\beta\simeq 0.3^\circ$, well above the requirement. This example clearly calls for innovation in the calibration strategy\cite{johnson/etal:2015,kaufman/keating/johnson:2016,nati/etal:2017,casas/etal:2021}.

When such precision is achieved for the angle calibration, we no longer have to rely on the Galactic foreground for measuring $\alpha$. The current measurement sets a requirement for precision: $\pm 0.05^\circ$ (i.e., $\pm 3'$) would yield a secure $>5\sigma$ detection of $\beta\simeq 0.3^\circ$. As the statistical uncertainty on $\beta$ expected from $EB$ data of future experiments is smaller\cite{pogosian/etal:2019}, the measurement will be limited by the calibration uncertainty.

Innovation is required not only for the calibration strategy, but also for the analysis technique. The techniques established for confirming the basic predictions of inflation - Gaussian statistics and nearly but not exactly scale-invariant power spectrum of scalar modes - should be developed for tensor modes\cite{shiraishi:2019,duivenvoorden/meerburg/freese:2020,hiramatsu/etal:2018,campeti/poletti/baccigalupi:2019}. Such tests are necessary for distinguishing between different origins of the primordial GW, i.e., quantum vacuum fluctuations in spacetime and GW sourced by matter fields during inflation. Equally important is the need for high-fidelity end-to-end simulations\cite{PlanckIntLVII}, with the $EB$ science and tests of Gaussianity in mind.

Last but not least, measuring the spectrum and polarisation of the stochastic background of GW across 21 decades in frequency (Fig.~\ref{fig:omega_gw}) sets an ambitious goal for ultimate tests of our ideas about the origin of the Universe. Such a research direction is in line with one of the three Large Mission science themes, `\textit{New Physical Probes of the Early Universe}', of the ESA Science Programme \textit{Voyage 2050}. Let's find new physics!


\section*{Acknowledgements}
This article is dedicated to the memory of Steven Weinberg.
We thank P. Campeti for sharing his work on the axion-SU(2) model, and J. Chluba, G. Dom\`enech, G. Dvali, J.~R. Eskilt, K. Lozanov, A. Maleknejad, Y. Minami, I. Obata, and M. Shiraishi for comments on the draft. We also thank anonymous referees for comments, which helped improve the presentation of the article.
The materials in this article are based partly on the Van der Waals Lecture delivered at the University of Amsterdam in 2020. We thank the institutes of Physics and Astronomy at the University of Amsterdam and the Vrije Universiteit Amsterdam for their hospitality, and the Stichting Van der Waals Fonds for the Johannes Diderik van der Waals rotating chair which enabled the visit.
This work was also supported in part by JSPS KAKENHI Grants No.~JP20H05850 and No.~JP20H05859, and the Deutsche Forschungsgemeinschaft (DFG, German Research Foundation) under Germany's Excellence Strategy - EXC-2094 - 390783311.
This work has also received funding from the European Union's Horizon 2020 research and innovation programme under the Marie Sk\l odowska-Curie grant agreement No.~101007633.
The Kavli IPMU is supported by World Premier International Research Center Initiative (WPI), MEXT, Japan.


\section*{Competing interests}
The author declares no competing interests.

\section*{Publisher’s note}
Springer Nature remains neutral with regard to jurisdictional claims in published maps and institutional affiliations.

\section*{Supplementary information}
\subsection*{Vacuum fluctuation in spacetime}
We quantise the tensor metric perturbation, $h_{ij}$, in vacuum. As the $+$ and $\times$ modes evolve independently in vacuum, we write $\Box h_p=0$ ($p=+,\times$). This equation takes the same form as the Klein-Gordon equation for a massless free field; thus, we can quantise $h_{p}$ following the standard procedure\cite{ford/parker:1977,birrell/davies:1984,mukhanov/feldman/brandenberger:1992}. We do not need full theory of quantum gravity here; we are only quantising \textit{fluctuations} in spacetime at linear order given the (classical) background spacetime.

With the conformal time $\eta$, the d'Alembert operator for a scalar field in the Friedmann-Robertson-Walker spacetime is given by $a^2\Box=-\partial^2/\partial \eta^2-2(a'/a)\partial/\partial \eta + \nabla^2$. Going to Fourier space, $h_{p}(\eta,\bm{x})=(2\pi)^{-3/2}\int d^3\bm{k}~h_{p,\bm{k}}(\eta)e^{i\bm{k}\cdot\bm{x}}$, and defining a convenient new variable, $u_{p,\bm{k}}(\eta)= C a(\eta)h_{p,\bm{k}}(\eta)$, where $C$ is a constant to be specified later, we write the equation of motion as\cite{lifshitz:1946,grishchuk:1975}
\begin{equation}
\label{eq:meff}
    u_{p,\bm{k}}'' + \left[k^2+m^2_\mathrm{eff}(\eta)\right]u_{p,\bm{k}}=0\,.
\end{equation}
Here $m^2_\mathrm{eff}(\eta)$ is a time-dependent effective mass-squared defined by $m^2_\mathrm{eff}\equiv -a''/a = -a^2H^2(2-\epsilon_H)$. Such a term did not appear in the equation of motion for $\bm{A}$ because of conformal invariance. Here we see that $u_p$ is not conformally invariant\cite{grishchuk:1975}.

As $\epsilon_H\ll 1$ during inflation, $m^2_\mathrm{eff}$ is actually \textit{negative}. The solution is divided into two regimes: (1) Super-Hubble regime ($k\ll |m_\mathrm{eff}| \simeq \sqrt{2}aH$) and (2) Sub-Hubble regime ($k\gg \sqrt{2}aH$). The former is a long-wavelength solution, whose physical wavelength is much greater than the Hubble length, $\lambda\gg H^{-1}$. The latter is a short-wavelength solution.

The sub-Hubble solution is oscillatory, $u_{p,\bm{k}}\propto e^{\pm ik\eta}$, so GW decays as $h_{p,\bm{k}}\propto a^{-1}e^{\pm ik\eta}$. The super-Hubble solution is $u_{p,\bm{k}}\propto a(\eta)$, hence $h_{p,\bm{k}}={\rm constant}$. This is remarkable: the primordial information is preserved in the super-Hubble regime. First, quantum-mechanical vacuum fluctuations are produced deep in the sub-Hubble regime. Then, exponential expansion stretches the wavelength of fluctuation to the super-Hubble scale, in which $h_{p,\bm{k}}$ becomes \textit{frozen}, preserving the primordial information of inflation. After inflation the Hubble length grows faster than the wavelength of GW, and GW comes inside the Hubble length, making it observable to us in CMB, motions of pulsars and stars, and interferometers.

If we ignore a small parameter $\epsilon_H$, the scale factor is given by $a(\eta)=-(H\eta)^{-1}$ for $-\infty<\eta<0$ and equation~(\ref{eq:meff}) can be solved exactly with a solution
\begin{align}
\nonumber
    u_{p,\bm{k}}(\eta)&=A_{p,\bm{k}}
    \left[\cos(k\eta)-\frac{\sin(k\eta)}{k\eta}\right]\\
    \label{eq:solution}
    &+ B_{p,\bm{k}}
    \left[\frac{\cos(k\eta)}{k\eta}+\sin(k\eta)\right]\,,
\end{align}
where $A_{p,\bm{k}}$ and $B_{p,\bm{k}}$ are integration constants. What determines them? Here comes quantum mechanics. Deep in the sub-Hubble regime, the effect of spacetime curvature is negligible. The sub-Hubble solution should therefore match the known flat-space solution for a free quantum field in vacuum called a `positive frequency mode'\cite{birrell/davies:1984}. Matching the two solutions, we find $A_{p,\bm{k}}$ and $B_{p,\bm{k}}$.

To this end, we must first find the correct variable (called `canonical variable') for quantisation. We wrote $u_{p}=Cah_{p}$, but what is $C$? To answer this, we expand the Einstein-Hilbert action for general theory of relativity, $I_\mathrm{GR}=(16\pi G)^{-1}\int d^4x\sqrt{-g}R$,
to second-order in tensor modes\cite{ford/parker:1977,mukhanov/feldman/brandenberger:1992}
\begin{equation}
    I^{(2)}_\mathrm{GR}=\int \frac{a^2d\eta d^3x}{16\pi G}\sum_{p=+,\times}\left[\frac12(h'_p)^2-\frac12(\nabla h_p)^2\right]\,.
\end{equation}
This is similar to the action for a free scalar field in Minkowski space, except the prefactor $a^2/(16\pi G)$. We define a canonically normalised field $u_p=ah_p/\sqrt{16\pi G}$, i.e., $C=1/\sqrt{16\pi G}$, to find the desired action
\begin{equation}
    I^{(2)}_\mathrm{GR}=\frac12\int d\eta d^3x\sum_{p}\left[(u'_p)^2-(\nabla u_p)^2-m^2_\mathrm{eff}(\eta)u_p^2\right]\,.
\end{equation}

We quantise $u_p$ using creation ($\hat a^\dagger_{\bm{k}}$) and annihilation ($\hat a_{\bm{k}}$) operators,
$u_{p,\bm{k}}(\eta)=u_{p,k}(\eta)\hat a_{p,\bm{k}}
+ u^*_{p,k}(\eta)\hat a^\dagger_{p,-\bm{k}}$, with the commutation relation $[\hat a_{p,\bm{k}},\hat a_{p',\bm{k}'}^\dagger]=\delta_{pp'}\delta_\mathrm{D}(\bm{k}-\bm{k}')$. A positive frequency mode is given by $u_{p,k}'=-i\omega_ku_{p,k}$ with an angular frequency $\omega_k$\cite{birrell/davies:1984}. We work in units of $\hslash=1$. The canonical commutation relation gives the normalisation condition $u_{p,k}u_{p,k}^*{}'-u^*_{p,k}u_{p,k}'=i$, which then gives $u_{p,k}=e^{-i\omega_k\eta}/\sqrt{2\omega_k}$.

We determine $A_{p,\bm{k}}$ and $B_{p,\bm{k}}$ in equation~(\ref{eq:solution}) as follows. Taking the sub-Hubble limit $k|\eta|\gg 1$ and comparing to the positive frequency solution, we find
$\omega_k=k$, $A_{p,\bm{k}}=(\hat a_{p,-\bm{k}}^\dagger+\hat a_{p,\bm{k}})/\sqrt{2k}$ and $B_{p,\bm{k}}=i(\hat a_{p,-\bm{k}}^\dagger-\hat a_{p,\bm{k}})/\sqrt{2k}$.
Finally, the full solution for $h_{p,\bm{k}}$ valid for all $k$ is
$h_{p,\bm{k}}(\eta)=h_{p,k}(\eta)\hat a_{p,\bm{k}}
+ h^*_{p,k}(\eta)\hat a^\dagger_{p,-\bm{k}}$ with
\begin{equation}
    h_{p,k} = i\frac{H}{M_{\mathrm{pl}}}\frac{e^{-ik\eta}}{k^{3/2}}\left(1+ik\eta\right)\,.
\end{equation}
The appearance of the reduced Planck mass, $M_{\mathrm{pl}}$, which contains $G$, $c$, and $\hslash$, is appropriate here because we quantised a linear perturbation of the gravitational field.

During inflation, the wavelength of quantum vacuum fluctuations is stretched to the super-Hubble scale, $k|\eta|\ll 1$, in which the solution approaches a constant, $h_{p,\bm{k}}\simeq i(H/M_{\mathrm{pl}})k^{-3/2}\hat c_{p,\bm{k}}$. The new operator $\hat c_{p,\bm{k}}$ is defined by $\hat c_{p,\bm{k}}\equiv \hat a_{p,\bm{k}}-\hat a_{p,-\bm{k}}^\dagger$, which commutes, $[\hat c_{p,\bm{k}},\hat c_{p,\bm{k}}^\dagger]=0$\cite{kiefer/polarski/starobinsky:1998,komatsu:2001}.
This poses an interesting question: does $h_{p,\bm{k}}$ remain a quantum field in the super-Hubble regime?
Of course, the full solution valid for all $k$ remains fully a quantum field, and we cannot take only a long-wavelength solution to claim that it has become a classical field.

No mechanism for `classicalisation' of a quantum field via, e.g., decoherence, was included in the above calculation. Rather, this is a manifestation of a squeezed quantum state in the super-Hubble regime\cite{grishchuk/sidorov:1989,grishchuk/sidorov:1990}: while it is a quantum field, it is indistinguishable from an ensemble of classical stochastic process\cite{polarski/starobinsky:1996,martin/vennin/peter:2012}. In other words, inflation creates gravitons, but their properties are very close to those of classical GW. Whether we can devise a way to probe quantum nature of the primordial GW remains an open question\cite{allen/flanagan/papa:2000,kanno/soda:2019}.
\subsection*{CMB polarisation from gravitational waves}
\begin{figure}[ht]
\centering
\includegraphics[width=\linewidth]{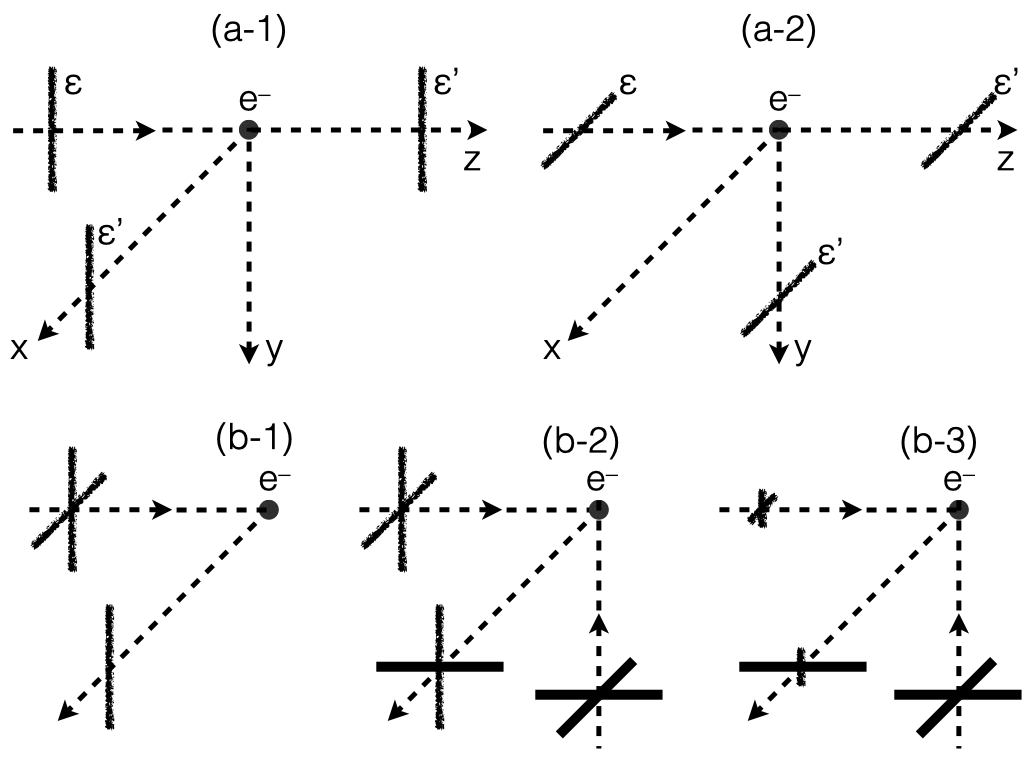}
\caption{{\bf Generation of polarisation.} An electron ($e^-$) at the origin receives an incident light along $z$ axis with the initial polarisation vector $\bm{\epsilon}$. The differential cross section is given by $d\sigma/d\Omega=3\sigma_\mathrm{T}(\bm{\epsilon}\cdot\bm{\epsilon}')^2/8\pi$, where $\sigma_\mathrm{T}$ is the Thomson scattering cross section and $\bm{\epsilon}'$ the polarisation vector of an outgoing light. {\bf a-1}| Incident light is vertically polarised and scattered into $x$ and $z$ axes. No scattering into $y$ axis is possible. {\bf a-2}| Incident light is horizontally polarised and scattered into $y$ and $z$ axes. No scattering into $x$ axis is possible. {\bf b-1}| Incident light is unpolarised, but light scattered into $x$ axis is vertically polarised. {\bf b-2}| Unpolarised lights are coming into $e^-$ along $y$ and $z$ axes with the equal intensity and scattered into $x$ axis. No polarisation is generated. {\bf b-3}| Finally, unpolarised lights are coming into $e^-$ along $y$ and $z$ axes with \textit{unequal} intensity and scattered into $x$ axis. Polarisation is generated.}
\label{fig:thomson}
\end{figure}

The degree of polarisation is proportional to a quadrupole moment of the local photon intensity distribution around electrons (Fig.~\ref{fig:thomson}). The observed polarisation depends also on the orientation of quadrupole with respect to observer's lines of sight. Polarisation pattern on the sky thus tells us the distribution of photon quadrupoles at the surface of last scattering\cite{hu/white:1997}.

Both the scalar and tensor modes generate a local quadrupole. In the tight-coupling regime before decoupling, in which photons and electrons are tightly coupled via Thomson scattering, the angular distribution of photon intensity around electrons is isotropic because electrons and photons move together on average - no polarisation is generated.
The electron number density fell exponentially as the temperature approached 3,000~K, which weakened coupling of photons and electrons and generated a quadrupolar angular distribution of photon intensity around electrons, hence polarisation.

\begin{figure}[ht]
\centering
\includegraphics[width=\linewidth]{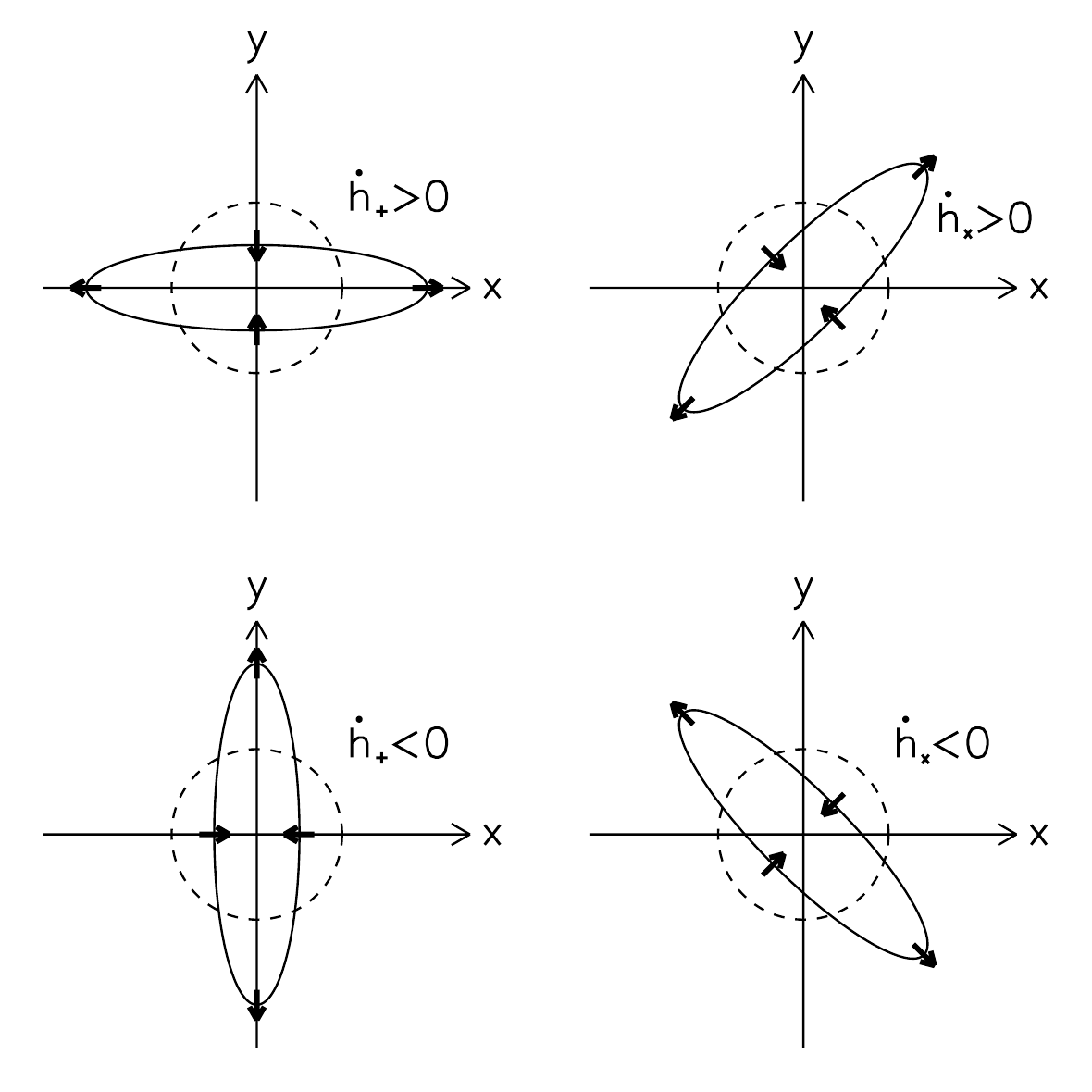}
\caption{{\bf Tensor-mode distortion of distances between two points.}}
\label{fig:gw}
\end{figure}

The tensor mode generates a quadrupole moment directly. In Fig.~\ref{fig:gw}, we show how $h_+$ and $h_\times$ propagating in the $z$ direction change distances between two points on $x$-$y$ plane. For example, when $h_+$ increases, the distance between two points increases along $x$ axis and decreases along $y$ axis such that the area is conserved.
Now imagine that space is filled with sea of photons. If space is stretched and contracted in $x$ and $y$ axes, wavelengths of photons are also stretched and contracted. This creates a quadrupole moment of photon intensity on the $x$-$y$ plane\cite{sachs/wolfe:1967}. Polarisation is generated when these photons are scattered by an electron at the origin of the plane into the $z$ direction towards an observer. Polarisation directions are parallel to the major axes of ellipses shown in Fig.~\ref{fig:gw}.

\begin{figure*}[ht]
\centering
\includegraphics[width=0.48\linewidth]{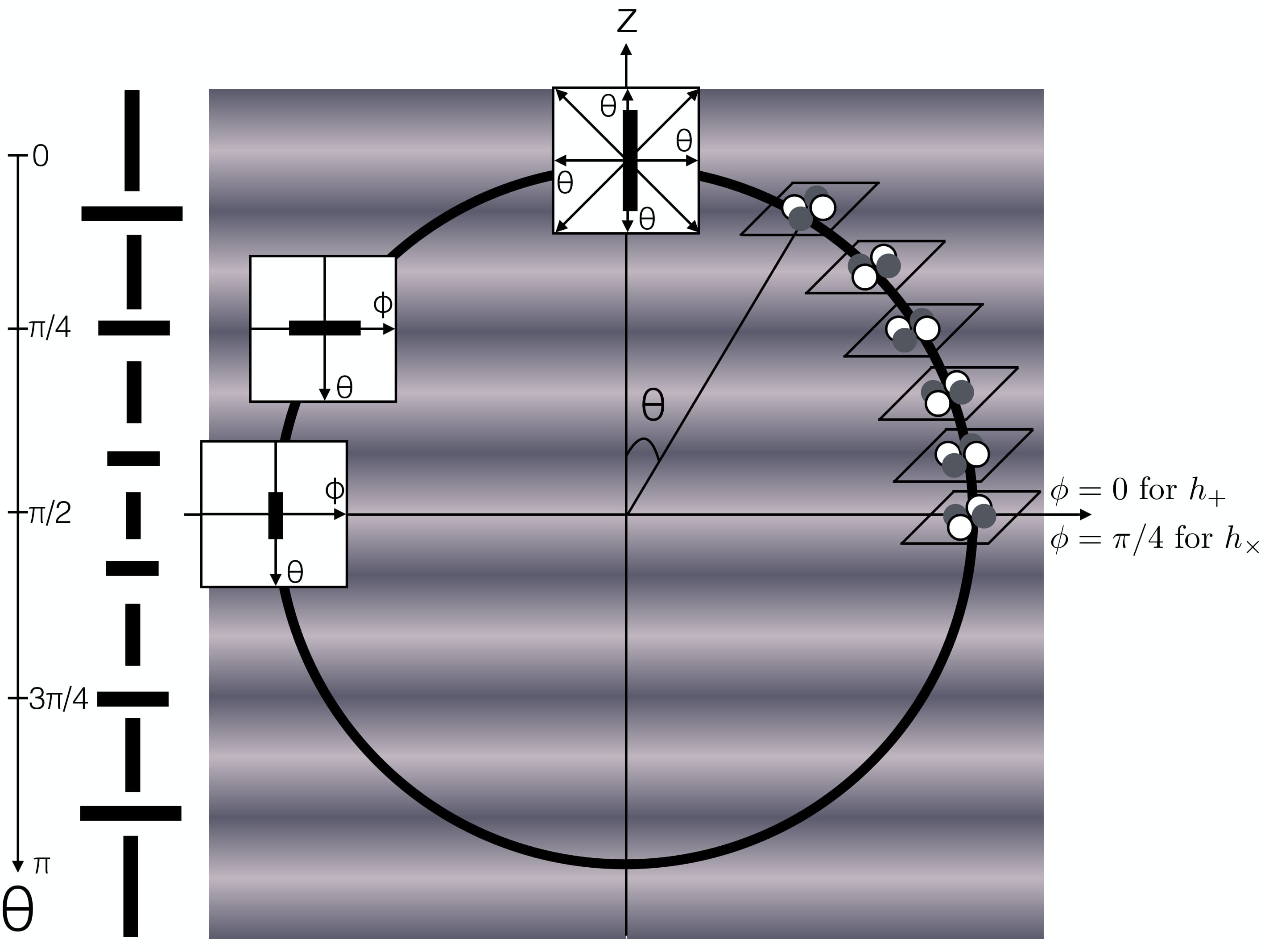}
\includegraphics[width=0.48\linewidth]{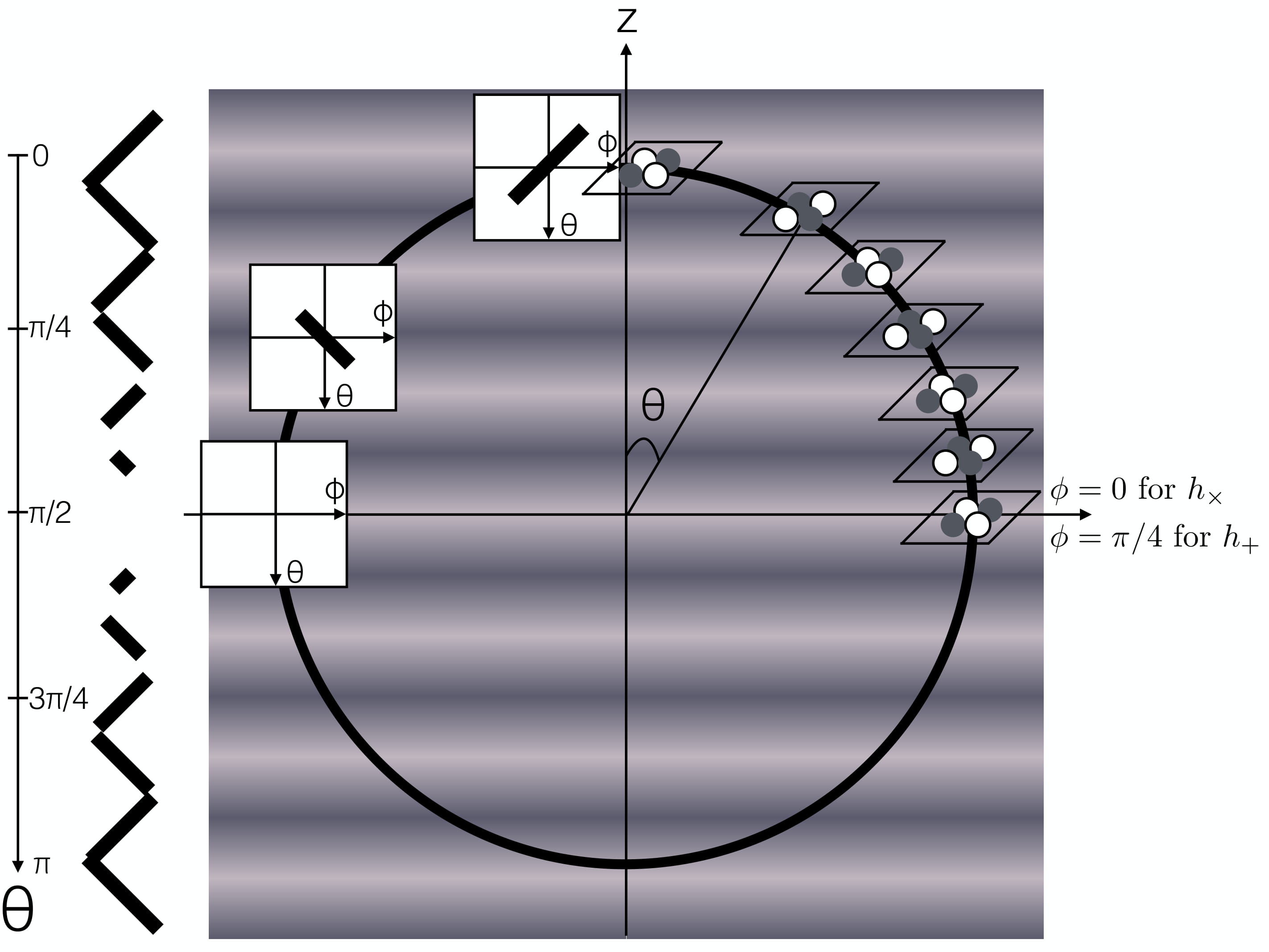}
\caption{{\bf Polarisation patterns of CMB on the sky generated by a single plane wave.}
When GW propagates in $z$ axis, quadrupole moments (illustrated by two grey and two white balls) of the photon intensity are generated on $x$-$y$ plane at each point on the surface of last scattering (solid circle). The white and grey balls show hot and cold photons as seen from electrons. When electrons scatter these quadrupoles, various strengths and directions of polarisation are seen by an observer located at the origin, depending on orientations of quadrupoles with respect to the observer's lines of sight. The left panel shows the polarisation pattern seen in azimuthal angles of $\phi=0^\circ$ and $45^\circ$ for $h_+$ and $h_\times$, respectively, whereas the right panel shows that of $\phi=45^\circ$ and $0^\circ$ for $h_+$ and $h_\times$, respectively.}
\label{fig:fourier}
\end{figure*}

The next step is to calculate the observed polarisation pattern on the sky. In Fig.~\ref{fig:fourier}, we show polarisation patterns in spherical coordinates $(\theta,\phi)$. The left panel shows the pattern we observe in $\phi=0^\circ$ ($45^\circ$), when a single plane wave of $h_+$ ($h_\times$) propagates in the $z$ axis. The right panel shows that in $\phi=45^\circ$ ($0^\circ$) for $h_+$ ($h_\times$).
The polarisation directions in the left panel are parallel or perpendicular to the propagating direction of GW (i.e., $E$ modes), whereas those in the right panel are tilted by $\pm 45^\circ$ ($B$ modes). Of course, fluctuations in our Universe are not described by a single plane wave but by a superposition of plane waves going in different directions with various wavelengths and amplitudes.

As the scalar mode does not produce $B$ modes at linear order in cosmological perturbation\cite{seljak:1997}, $B$ modes are a clean probe of the primordial GW\cite{seljak/zaldarriaga:1997,kamionkowski/etal:1997b}.

\bibliography{references}
\end{document}